\documentclass[fleqn,usenatbib]{mnras}

\usepackage{newtxtext,newtxmath}
\usepackage{soul}
\usepackage{natbib}
\usepackage{pdflscape}
\usepackage[T1]{fontenc}
\usepackage{rotating}
\usepackage{lmodern}
\DeclareRobustCommand{\ion}[2]{%
\relax\ifmmode
\ifx\testbx\f@series
{\mathbf{#1\mathsc{#2}}}\else
{\mathrm{#1\mathsc{#2}}}\fi
\else\textup{#1{\mdseries\textsc{#2}}}%
\fi}

\newcommand{\Hii}{\ion{H}{II}}

\newcommand{\Halpha}{\ion{H}{$\alpha$}}
\newcommand{\Hbeta}{\ion{H}{$\beta$}}
\newcommand{\oiii}{[\ion{O}{iii}]}

\newcommand{\sii}{[\ion{S}{ii}]}
\newcommand{\nii}{[\ion{N}{ii}]}

\usepackage{graphicx}	
\usepackage{amsmath}	

\usepackage{subcaption}
\usepackage{threeparttable} 
\usepackage{booktabs}

\usepackage{orcidlink}

\title[SDSS-V LVM: Structure of the Rosette Nebula]{SDSS-V Local Volume Mapper (LVM): Revealing the Structure of the Rosette Nebula}
\author[M. A. Villa-Durango et al.]{
Mónica A. Villa-Durango\,\orcidlink{0009-0008-1605-4771}$^{1}$\thanks{E-mail: mavillad@astro.unam.mx}, 
 Jorge Barrera-Ballesteros\,\orcidlink{0000-0003-2405-7258}$^{2}$, 
 Carlos G. Román-Zúñiga\,\orcidlink{0000-0001-8600-4798}$^{1}$, \and
 Emma R. Moran\,\orcidlink{0009-0002-1352-8296}$^{3}$,
Jason E. Ybarra\,\orcidlink{0000-0002-3576-4508}$^{4}$, 
 J. Eduardo Méndez-Delgado\,\orcidlink{0000-0002-6972-6411}$^{2}$, 
 Niv Drory \,\orcidlink{0000-0002-7339-3170}$^{5}$, \and
 Kathryn Kreckel\,\orcidlink{0000-0001-6551-3091}$^{6}$, 
Hector Ibarra-Medel \,\orcidlink{0000-0002-9790-6313}$^{2}$,
 S. F. Sánchez\,\orcidlink{0000-0001-6444-9307}$^{1,7}$,
 Evelyn J. Johnston\,\orcidlink{0000-0002-2368-6469}$^{8}$, \and
 A. Roman-Lopes\,\orcidlink{0000-0002-1379-4204}$^{9}$, 
Jes\'us Hernandez\,\orcidlink{0000-0001-9797-5661}$^{1}$,
 Jos\'e G. Fern\'andez-Trincado\,\orcidlink{0000-0003-3526-5052}$^{10}$, 
 Amelia M.\ Stutz \,\orcidlink{0000-0003-2300-8200}$^{11}$, \and
 William J. Henney\,\orcidlink{0000-0001-6208-9109}$^{12}$,
A. Ghosh\,\orcidlink{0000-0001-7650-1870}$^{12}$,
 Sumit K. Sarbadhicary\,\orcidlink{0000-0002-4781-7291}$^{13}$, 
 A. Z. Lugo-Aranda\,\orcidlink{0000-0001-9226-9178}$^{1}$,\and 
 Dmitry Bizyaev\,\orcidlink{0000-0002-3601-133X}$^{14,15}$,
 Amy M.\ Jones\,\orcidlink{0000-0002-2262-8240}$^{16}$,
Guillermo A. Blanc\,\orcidlink{0000-0003-4218-3944}$^{17,18}$
\\
$^{1}$ Universidad Nacional Autónoma de México, Instituto de Astronomía, AP 106, Ensenada, 22800, BC, Mexico; \\
$^{2}$ Universidad Nacional Autónoma de México, Instituto de Astronomía, AP 70-264, CDMX 04510, Mexico;\\
$^{3}$ Department of Physics \& Astronomy, University of Pittsburgh, Pittsburgh, PA 15260, USA;\\
$^{4}$ Department of Physics, Davidson College, 405 N. Main St., Davidson, NC 28035, USA;\\
$^{5}$ McDonald Observatory, The University of Texas at Austin, 1 University Station, Austin, TX 78712-0259, USA;\\
$^{6}$ Astronomisches Rechen-Institut, Zentrum f\"ur Astronomie der Universit\"at Heidelberg, M\"onchhofstra\ss e 12-14, D-69120 Heidelberg, Germany;\\
$^{7}$ Instituto de Astrofísica de Canarias, La Laguna, Tenerife, Spain, E-38200\\
$^{8}$ Instituto de Estudios Astrof\'isicos, Facultad de Ingenier\'ia y Ciencias, Universidad Diego Portales, Av. Ej\'ercito Libertador 441, Santiago, Chile;\\
$^{9}$ Department of Astronomy, Universidad de La Serena, Av. Raul Bitran 1302, La Serena, Chile;\\
$^{10}$ Instituto de Astronom\'ia, Universidad Cat\'olica del Norte, Av. Angamos 0610, Antofagasta, Chile;\\
$^{11}$ Departamento de Astronom\'{i}a, Universidad de Concepci\'{o}n,Casilla 160-C, Concepci\'{o}n, Chile\\
$^{12}$ Instituto de Radioastronomía y Astrofísica (IRyA), UNAM Campus Morelia, Apartado postal 3-72, 58090 Morelia, Michoacan, Mexico;\\
$^{13}$ Department of Physics and Astronomy, The Johns Hopkins University, Baltimore, MD 21218 USA;\\
$^{14}$ Apache Point Observatory and New Mexico State University, P.O. Box 59, Sunspot, NM, 88349-0059, USA;\\
$^{15}$ Sternberg Astronomical Institute, Moscow State University, Moscow;\\
$^{16}$ Space Telescope Science Institute, 3700 San Martin Drive, Baltimore, MD 21218, USA;\\
$^{17}$ Observatories of the Carnegie Institution for Science, 813 Santa Barbara Street, Pasadena, CA 91101, USA ;\\
$^{18}$ Departamento de Astronom\'{i}a, Universidad de Chile, Camino del Observatorio 1515, Las Condes, Santiago, Chile
}

\date{Accepted XXX. Received YYY; in original form ZZZ}

\pubyear{2025}

\begin{document}
\label{firstpage}
\pagerange{\pageref{firstpage}--\pageref{lastpage}}
\maketitle

\begin{abstract}
The Rosette Nebula is a well-known H II region shaped by the interaction of gas with the OB stars of the NGC 2244 stellar association. Located within the remnant of a giant molecular cloud, it exhibits a complex structure of ionized gas, molecular material, dust, and embedded clusters. In October 2023, the region was observed as part of the SDSS-V Local Volume Mapper (LVM) integral field spectroscopy survey. Covering a radius of $\sim$1°, the dataset comprises 33,326 spectra with spatially resolved information spanning 390–980 nm.
We present a structural analysis of the ionized, molecular, and dusty components using multi-wavelength observations: optical spectroscopy from SDSS-V LVM, $^{12}$CO emission from PMO/MWISP (sub-millimeter), and dust emission from WISE (\(12 \, \mu\text{m}\)) and Herschel (far-infrared). These datasets were complemented with the positions of ionizing stars to study emission structures traced by \Halpha, \Hbeta, \oiii, \nii, and \sii\, as well as the spatial distribution of line ratios (\Halpha/\Hbeta, \oiii/\Hbeta, \nii/\Halpha, and \sii/\Halpha) relative to the surrounding molecular cloud.
Our analysis reveals interaction zones between ionized and neutral gas, including filaments, globules, and dense regions with or without ongoing star formation. Radial and quadrant-based flux profiles further highlight morphological and ionization variations, supporting the scenario in which the Rosette Nebula evolved from a non-homogeneous molecular cloud with a thin, sheet-like structure.

\end{abstract}

\begin{keywords}
ISM: clouds - 
HII regions - 
ISM: lines and bands -
ISM: structure.
\end{keywords}



\section{Introduction}

\Hii\ regions are fundamental to galactic evolution, as they trace key processes such as star formation, interstellar material recycling, and chemical enrichment of the interstellar medium, \cite{Fabian2023}. These regions, associated with young OB stars formed in stellar clusters within molecular clouds, are shaped by intense ultraviolet radiation and stellar winds, which ionize and disperse the surrounding gas, altering the structure of the ISM and potentially triggering new episodes of star formation \cite[e.g.][]{Hopkins2014, feedback2014, McLeod2015}. A detailed study of the gas morphology and kinematics in these regions, through spatially resolved spectroscopic observations, provides insights into how stellar feedback regulates the evolution of star-forming complexes \citep{Kim2018, Barrera2021, McLeod2019}.

The Local Volume Mapper (LVM; \citealp{Niv2024}), one of three programs in the fifth generation of the Sloan Digital Sky Survey (SDSS-V; \citealp{Kollmeier2025}), is an integral field spectroscopic (IFS) survey designed to map star-forming regions in the Milky Way, the Magellanic Clouds, and galaxies of the local-volume. The primary goal of LVM is to understand the interaction between stellar feedback and the interstellar medium (ISM), on both small and large scales. The survey aims to spatially resolve the physical properties of the ISM, including its kinematics, particularly in star-forming regions. This will allow to investigate the physical processes that regulate star formation and ISM dynamics. Covering more than 4,300 square degrees, LVM will collect over 55 million spectra, providing an unprecedented spatial resolution ranging from 0.05 to 1 parsec in the Milky Way, 10 parsecs in the Magellanic Clouds, and larger scales in nearby galaxies.

One of the main scientific goals of the LVM is to obtain IFS data across hundreds of Galactic \Hii\ regions, many of them extended enough to allow LVM to study them in exquisite detail. For example, the LVM Orion Nebula project \citep{Kreckel2024} described the data collected from 108 tiles covering a radius of 2 degrees in the sky, revealing flux maps based on the widely used forbidden collisional lines \oiii, \nii, and \sii, and recombination emission lines like \Halpha\ and \Hbeta. Such analysis enabled the exploration of the ISM around some well known ionizing sources. It revealed the structure of the interface with the molecular cloud, highlighting the rich scientific potential of LVM for studying \Hii\ regions in the Milky Way.

Another region selected as an Early Science observation was the Rosette Nebula (RN), an archetypal \Hii\ region. At its center, the RN hosts the OB star association of the NGC 2244 cluster, which is responsible for the ionization of the region. This cluster is relatively young, with an estimated age of approximately 2 Myr \citep{Lim2021}. Several studies place the RN at distances ranging between 1.3 and 1.8 kpc (e.g., \cite{ReviewRoman2008} and references therein). Recent studies have constrained the distance to RN to lie between 1.4 and 1.5 kpc. For example, \citet{Lim2021} adopt a value of 1.4 kpc based on Gaia parallaxes of stars in the NGC 2244 cluster, while \citet{Muzic2022} assume a distance of 1.5 kpc, derived from a comparative analysis of the distances to both NGC 2244 and NGC 2237. In this work, we adopt a heliocentric distance of 1.5 kpc. 

Over the past decades, various studies have sought to understand how the interaction between OB stars and the surrounding medium has shaped the central cavity and the ionized gas structure observed in the RN. For example, \cite{Bruhweiler2010} analyzed the central cavity formed by the stellar bubble generated by the winds from the O-type stars of NGC 2244. Their results indicate that the hot bubble at the center of the nebula is expanding at 56 km/s, but its size and age are significantly smaller than theoretical predictions. They estimated that the bubble's age is approximately 64,000 years, a value considerably lower than the estimated age of the cluster. On the other hand, \cite{Wareing2018} developed hydrodynamical and magnetohydrodynamical models to investigate the evolution and morphology of the Rosette Nebula. Their simulations indicate that the stellar wind from HD 46150, the most massive O-type star at the center of the nebula, has been the primary driver of the formation and evolution of the central cavity. 
Moreover, they demonstrated that if the progenitor molecular cloud had a thin-sheet structure, resulting from interactions with a perpendicular magnetic field, the observed RN structure can be reproduced through stellar feedback mechanisms. In this model, the structure of the nebula is successfully reproduced on a timescale of approximately $\sim$1.73 Myr, which is more consistent with the estimated age of the central cluster, thereby resolving the previously noted discrepancy between the dynamical age of the nebula and the cluster age

The RN is part of a larger complex that includes an adjacent molecular cloud where a collection of 11 new stellar clusters are forming \citep{Cambresy2013}. Consequently, some studies have focused on the evolution of embedded stellar clusters in the region. Through infrared and X-ray studies, stellar groupings have been identified in interface zones between the \Hii\ region and the remaining molecular cloud \citep[e. g. NGC 2237, PL02]{PL97,Roman2008}. 
Additional efforts have focused on analyzing high-density areas at the ionization front and identifying filaments or globules that may promote the formation of new stellar objects \citep{Gahm2007, Wang2010, Makela2014}. This makes the RN an ideal setting to study the interaction between OB stars and ionized gas, as well as the role these interactions may play in promoting or inhibiting  star formation \citep{Hopkins2014, feedback2014, Muzic2019}. Therefore, the morphological study of the different components of the ISM in this region is essential to better understand these interactions and their implications for the evolution of the nebula.

The importance of the RN as a target lies in two main aspects. The first is the layout of the central ionized region, which contains relatively little dust and offers a clear view of the central photodissociation region (PDR), making it favorable for studying the OB association and ionized gas. The second is the structure of the surrounding \Hii\ region, which is highly symmetric and thus ideal for investigating radial variations and gradients. In addition, the RN has been extensively studied across multiple wavelengths, providing not only a rich body of literature but also a valuable dataset for comparing physical processes in the ISM \cite[e.g.][]{Gahm2007,Makela2017,Ybarra13}. In particular, the spatial distribution of molecular gas has been mapped using millimeter-wavelength emission from carbon monoxide (CO) isotopes. \cite{Dent2009} explored the kinematics of molecular structures in the Rosette Nebula and the influence of OB stars with young clusters in their local environment. They analyzed the molecular gas dynamics in response to the radiation from O-type stars, particularly in high-density regions or globules. They found that these globules tend to exhibit velocity gradients, indicating motion away from the center of the NGC 2244 cluster. Additionally, they found evidence of an expanding molecular ring with a radius of 11 pc.

In this article, we focus on a morphological description of the RN to understand its ionization structure through optical data obtained with the LVM, exploring the relationship between different ISM components, similar to the analysis performed for the Orion Nebula dataset \citep{Kreckel2024}. The LVM data enable direct comparisons with other components of the interstellar medium and the associated massive stars. This study represents an initial step toward understanding the ionization physics in the region and lays the foundation for future kinematic analyses of its various components, which will be presented in a forthcoming article.

The aim of this work is to provide a morphological analysis of the RN using multiwavelength data, with a spectral focus on optical emission line maps from the LVM dataset. In Section 2, we describe the various surveys and data used in the analysis; in Section 3, we examine the LVM data, presenting maps of different emission lines and a comparison between the ionization region and the remaining molecular cloud. In Section 4, we discuss some areas of interest that underscore the role of OB complex stars in shaping the region’s morphology. Finally, conclusions are presented in Section 5.

\section{Data and Analysis}
\label{sec:data}
\subsection{Local Volume Mapper}

\begin{figure*}
    \centering
    \includegraphics[width=0.95\textwidth]{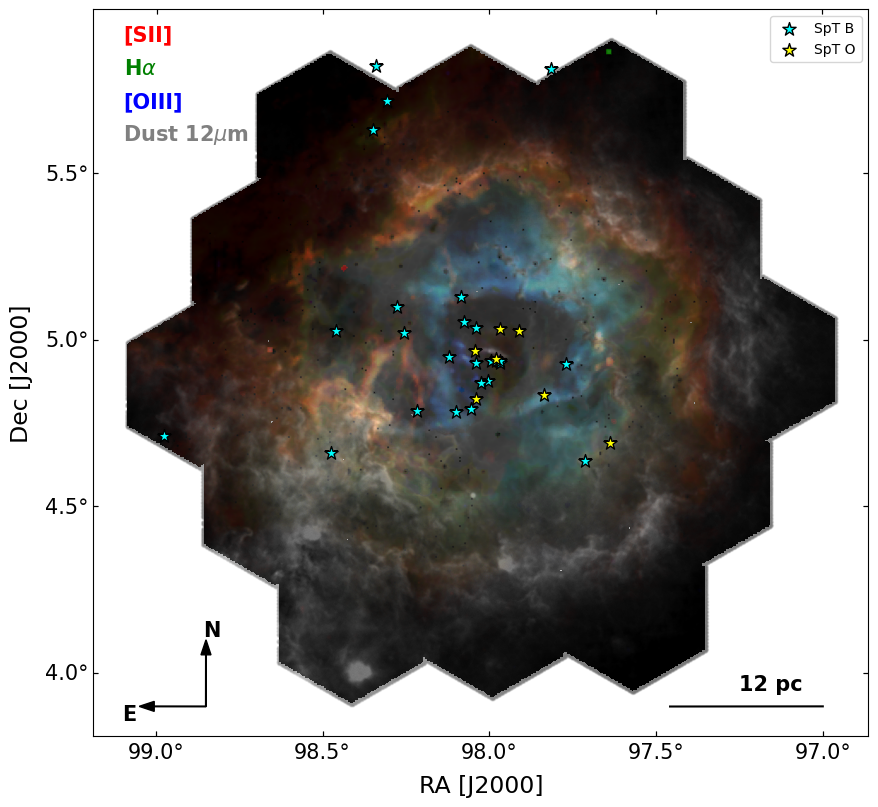} 
    \caption{Recreated image of the Rosette Nebula using LVM data and WISE dust maps. The image reveals many ionized gas structures and the surrounding dust distribution.
    The emission-line maps of \oiii\ (blue), \Halpha\ (green), and \sii\ (red) are displayed, while the \(12 \, \mu\text{m}\) dust emission is shown in grayscale (units of MJy sr$^{-1}$). O- and B-type stars are marked in yellow and cyan, respectively.
    }
    \label{fig:Rosetta-RGB}
\end{figure*}

The LVM instrument \citep[LVM-I][]{Konidaris2024} consists of four 16 cm diameter telescopes fitted with lenslet-coupled fiber systems, and spectrographs that cover a wavelength range of 3600-9800 \(\text{\r{A}}\) at a spectral resolution of approximately R $\approx$ 4000 \citep{Perruchot2018}. Three of the four telescopes are used for observing sky spectra and spectrophotometric standard stars. The science telescope (Sci) is fitted with a hexagonal IFU that has a $\sim$0.5 degrees diameter field of view. The IFU is sampled by 1801 lenslet-coupled optical fibers, each with an angular diameter of 35.3 arcseconds on the sky. The telescopes are installed at Las Campanas Observatory, Chile \citep{Herbst2024}.

The LVM Data Reduction Pipeline (DRP) processes the large volume of spectral data by extracting, calibrating, and correcting for instrumental effects such as cosmic rays and sky background \citep{mejia24}. Subsequently, these data are analyzed by the Data Analysis Pipeline (DAP), which decomposes the observed spectra into stellar and nebular components, retrieving key physical properties such as emission line fluxes, velocity dispersions, and equivalent widths \citep{Sanchez2025}.
However, at the time of writing this article, the DRP is under active development; while it provides an absolute flux calibration, this calibration is preliminary and is being refined. Consequently, the analysis presented in this study is based on relative fluxes, and we focus on qualitative rather than quantitative comparisons of the emission.

Observations for the RN were taken during the early science commissioning of LVM, on the nights of 4, 10, and 13 October 2023(MJD 60222, 60228 and 60231). These observations consisted of 19 tiles, covering the entire optical extent of the nebula (see Fig. \ref{fig:Rosetta-RGB}). After the reduction process, a total of 33,356 spectra were obtained. For this study, we use the integrated flux of the emission lines \Halpha\ $\lambda$ 6563\(\text{\r{A}}\), \Hbeta\ $\lambda$ 4861\(\text{\r{A}}\), \oiii\ $\lambda$ 5007\(\text{\r{A}}\), \nii\ $\lambda$ 6583\(\text{\r{A}}\), and \sii\ $\lambda \lambda$ 6717\(\text{\r{A}}\), 6731\(\text{\r{A}}\). 
For these emission lines, only data with a signal-to-noise ratio (S/N) greater than 1 were used, and for \Halpha\ specifically, S/N $\geq$ 10 (23,987 spectra in total). The fluxes obtained with the DAP for these spectra were divided by an arbitrary value of $10 \times 10^{-16}$ erg s$^{-1}$ cm$^{-2}$ to obtain relative fluxes.

At the adopted heliocentric distance of 1.5 kpc, the $\sim$2 degree of the field of view of the observations corresponds to approximately 52 pc on the sky. This implies that the nebula spans a projected radius of $\sim$ 27 pc, and each LVM fiber spans an aperture of $\sim$0.26 pc. If a distance of 1.4 kpc were assumed, these values would decrease slightly to $\sim$ 49 pc in total extent, $\sim$25 pc in radius, and $\sim$0.24 pc per fiber. This difference in the adopted distance does not significantly affect the analysis and conclusions presented in this work.

We present a composite color image in Fig. \ref{fig:Rosetta-RGB}, combining \oiii, \Halpha, \sii, plus \(12 \, \mu\text{m}\) dust emission in grayscale, to highlight morphological structure. This composition highlights the archetypal structure of an \Hii\ region bounded by the remnants of the molecular cloud. The evacuated central cavity is clearly visible, surrounded by ionized gas and hot dust, heated by the intense radiation from the massive stars in the cluster.

\subsection{MWISP millimeter emission maps}

A more recent and comprehensive view of the Rosette Molecular Cloud (RMC) has been provided by the Milky Way Imaging Scroll Painting (MWISP\footnote{http://www.radioast.nsdc.cn/mwisp.php}; \citealp{mwisp2018}) survey, which conducted large-scale mapping of the $\mathrm{^{12}CO}$, $\mathrm{^{13}CO}$, and $\mathrm{C^{18}O}$ (J=1–0) emissions using the Purple Mountain Observatory (PMO) 13.7-meter millimeter telescope. This study enabled the determination of the temperature, velocity, and dispersion of the molecular cloud, as well as improved our understanding of its structure, dynamics, and the influence of stellar clusters. It provides a detailed view of the interaction between stellar radiation and the molecular cloud.

For this project, we make use of the $\mathrm{^{12}CO}$ line emission maps of the RN from MWISP. The map covers an area of $\mathrm{3.5^\circ\times 2.5^\circ}$ centered at ($l,b$)=(206.5,-2.25). The map has a nominal spectral resolution of $\mathrm{0.16\ km\cdot s^{-1}}$ at the central frequency of 115 GHz, with a pointing accuracy of 5$\arcsec$, and beamwidths of 55 and 52$\arcsec$ at 110 and 115 GHz, respectively. The spectral cubes we use are FITS cubes with a pixel grid of 30$\arcsec$, and 1387 channels covering a radial velocity interval of $\mathrm{-10\ to\ 13\ km\cdot\ s^{-1}}$ This CO map provides crucial information about the spatial and kinematic distribution of molecular gas in the Complex. 

\subsection{Thermal dust emission map}

The Wide-field Infrared Survey Explorer (WISE; \citealt{Wright2010}) conducted an all-sky survey in four mid-infrared bands (3.6, 4.5, 12, and 22 $\mu$m) with an angular resolution of 6$\arcsec$. 
The thermal dust emission at 12 $\mu$m was mapped as part of the WISE Sky Survey Atlas (WSSA; \citealp{meisner2014}), and an effort to reprocess the images obtained by the W3 band, removing as many possible artifacts (from point sources to solar system object trails) to produce a Galactic dust emission map. The WSSA survey is divided in 430 tiles, 156.25 $(\deg)^2$ each, zero-point calibrated and smoothed to 15$\arcsec$. We used the \texttt{Montage}\footnote{http://montage.ipac.caltech.edu/} package to combine tiles 208 and 244 of the survey, both of which cover the Rosette Complex. The resulting map traces, in great detail, the thermal emission from UV photon and cosmic-ray heated polycyclic aromatic hydrocarbon (PAHs) molecules across the reflection nebula and the adjacent Giant Molecular Cloud. These observations complement molecular cloud studies by providing key information on the evolution of molecular globules and the photoevaporation of gas at the periphery of the \Hii\ region.

\subsection{Herschel column density map}

As part of the HOBYS (Herschel Imaging Survey of OB Young Stellar Objects, \cite{Motte2010}) program, the Herschel Space Observatory provided a map of the Rosette Molecular Cloud, including the southeastern region of RN \citep{Schneider10}. Emission data in the far-infrared (70–500 $\mu$m) were obtained using the SPIRE (\cite{Griffin2010, Swinyard2010}) and PACS (\cite{Poglitsch2010}) instruments. Based on the far-infrared maps of the Rosette Molecular Cloud carried out by the Herschel Space Observatory, \cite{Moran+2024} constructed a column density map covering the southeastern part of the RN. This map was obtained by fitting the dust temperature at each position of the maps, using spectral energy distributions constructed from individual measurements at 70 and 160 $\mu$m from PACS, and 230 and 350 $\mu$m from SPIRE. From those fits it is possible to derive the opacity of the dust which can be then converted to dust extinction and column density.

\subsection{Stars OB catalog}

To conduct a detailed analysis of the morphology and structure of the ISM with respect to the location of the ionizing stars, we identified stars classified as spectral type(SpT) O and B within the field of view of the LVM pointings. For the O-type stars, we performed a literature review focused on the RN, identifying a total of seven stars belonging to the NGC 2244 cluster, as cataloged by \cite{Wang2008, Mahy2009, Martins2012} and \cite{Lim2021}. For the selection of B-type stars, we used the SIMBAD database, applying a distance filter between 1.3 and 1.7 kpc, based on the study by \cite{Muzic2022}. Using this criterion, we selected stars with spectral types between B0 and B5. The data for these stars are presented in Table 1 (O-SpT stars) and Table 2 (B-SpT stars).The distances listed correspond to the \textit{rgeo} values from the Gaia DR3 catalog as estimated by \citet{BailerJones2021}

\begin{table}
    \centering
    \caption{NGC 2244 O-type stars in the Rosette Nebula field.}
    \label{tab:Stars_O}
    \begin{threeparttable}
    \renewcommand{\arraystretch}{1.5}
    \resizebox{\columnwidth}{!}{
    \begin{tabular}{l|c|c|c|c}
        \toprule
        Stars  & RA [deg] & DEC [deg] & SpT$^{5}$ & Dist [kpc] \\ 
        \midrule
         HD 258691 \tnote{1,3}      & 97.6387  & 4.6910  & O9.5V      & 1.617  $^{\text{+0.290}}_{\text{--0.272}}$ \\ 
         HD 46056 \tnote{1,2,3,4}  & 97.8369  & 4.8344  & O8Vn        & 1.412	 $^{\text{+0.046}}_{\text{--0.046}}$ \\ 
         HD 46223 \tnote{1,2,3}     & 98.0388  & 4.8235  & O4V((f))   & 1.390	 $^{\text{+0.048}}_{\text{--0.056}}$ \\ 
         HD 46202 \tnote{1,2,3}     & 98.0436  & 4.9666  & O9.2V      & 1.488	 $^{\text{+0.084}}_{\text{--0.065}}$ \\ 
         HD 46150 \tnote{1,2,3,4}   & 97.9813  & 4.9429  & O5V((f))z  & 1.473	 $^{\text{+0.036}}_{\text{--0.203}}$ \\ 
         HD 46106 \tnote{4}         & 97.9100  & 5.0268 & O9.7III(n) & 1.490	 $^{\text{+0.121}}_{\text{--0.079}}$ \\ 
         HD 46149 \tnote{1,2,3,4}   & 97.9689  & 5.0331  & O8.5V      & 1.485	 $^{\text{+0.073}}_{\text{--0.058}}$ \\  
        \bottomrule
    \end{tabular}
    }
    \begin{tablenotes}
    \item \parbox{\textwidth}{\footnotesize 
    [1] \cite{Wang2008}, [2] \cite{Mahy2009}, [3] \cite{Martins2012},} 
    \item \parbox{\textwidth}{\footnotesize
    [4] \cite{Lim2021}, [5] Simbad.}
    \end{tablenotes}
    
    \end{threeparttable}
\end{table}

\begin{table}
    \centering
    \caption{NGC 2244 B-type stars in the Rosette Nebula field.}
    \label{tab:Stars_B}
    \begin{threeparttable}
    \renewcommand{\arraystretch}{1.5}
    \resizebox{\columnwidth}{!}{
    \begin{tabular}{l c c c c}
        \toprule
        Stars              & RA [deg] & DEC [deg] & SpT & Dist [kpc] \\ 
        \midrule
        V* V578 Mon        & 98.0025   & 4.8780     & B0V+B1V  & 1.358 $^{\text{+0.045}}_{\text{--0.039}}$ \\  
        NGC 2244 190       & 97.9954   & 4.9378     & B2.5Vn   & 1.517 $^{\text{+0.052}}_{\text{--0.058}}$ \\  
        LS VI +04 12       & 98.0256   & 4.8709     & B1III    & 1.480 $^{\text{+0.122}}_{\text{--0.116}}$ \\  
        NGC 2244 123       & 97.9801   & 4.9394     & B5V      & 1.430 $^{\text{+0.044}}_{\text{--0.062}}$ \\  
        HD 259105          & 97.9667   & 4.9326     & B1V      & 1.388 $^{\text{+0.038}}_{\text{--0.032}}$ \\  
        NGC 2244 125       & 97.9688   & 4.9375     & B4V      & 1.393 $^{\text{+0.031}}_{\text{--0.036}}$ \\  
        NGC 2244 192       & 98.0400   & 4.9324     & B3/5     & 1.376 $^{\text{+0.035}}_{\text{--0.041}}$ \\  
        NGC 2244 206       & 98.0560   & 4.7936     & B4V      & 1.396 $^{\text{+0.056}}_{\text{--0.060}}$ \\  
        HD 259300          & 98.1224   & 4.9489     & B3Vp     & 1.491 $^{\text{+0.087}}_{\text{--0.088}}$ \\  
        LS VI +05 8        & 98.0410   & 5.0371     & B2.5V    & 1.470 $^{\text{+0.074}}_{\text{--0.077}}$ \\  
        LS VI +04 14       & 98.1010   & 4.7844     & B2.5V    & 1.349 $^{\text{+0.044}}_{\text{--0.047}}$ \\  
        HD 259238          & 98.0759   & 5.0560     & B0V      & 1.395 $^{\text{+0.040}}_{\text{--0.055}}$ \\  
        NGC 2244 36        & 97.7690   & 4.9287     & B5       & 1.319 $^{\text{+0.036}}_{\text{--0.037}}$ \\  
        NGC 2244 334       & 98.2158   & 4.7878     & B5V      & 1.369 $^{\text{+0.037}}_{\text{--0.038}}$ \\  
        NGC 2244 231       & 98.0854   & 5.1314     & B5       & 1.398 $^{\text{+0.033}}_{\text{--0.039}}$ \\  
        NGC 2244 351       & 98.2566   & 5.0218     & B5       & 1.470 $^{\text{+0.049}}_{\text{--0.052}}$ \\  
        NGC 2244 345       & 98.2774   & 5.1010     & B2       & 1.513 $^{\text{+0.055}}_{\text{--0.073}}$ \\  
        LS VI +05 12       & 98.4607   & 5.0271     & B2.5V    & 1.459 $^{\text{+0.037}}_{\text{--0.054}}$ \\  
        HD 46484           & 98.4767   & 4.6624     & B2Ib     & 1.497 $^{\text{+0.034}}_{\text{--0.035}}$ \\  
        TYC 158-2924-1     & 98.3511   & 5.6324     & B3       & 1.445 $^{\text{+0.033}}_{\text{--0.026}}$ \\  
        HD 259508          & 98.3070   & 5.7183     & B5       & 1.494 $^{\text{+0.056}}_{\text{--0.044}}$ \\  
        GSC 00158-00640    & 97.8130   & 5.8143     & B5       & 1.470 $^{\text{+0.045}}_{\text{--0.036}}$ \\  
        GSC 00158-00500    & 98.3396   & 5.8241     & B5       & 1.579 $^{\text{+0.046}}_{\text{--0.057}}$ \\  
        GSC 00154-00818    & 98.9764   & 4.7108     & B4V      & 1.463 $^{\text{+0.035}}_{\text{--0.035}}$ \\  
        Cl* NGC 2244 CDZ 36& 97.7117   & 4.6367     & B5       & 1.356 $^{\text{+0.029}}_{\text{--0.026}}$ \\   
        \bottomrule
    \end{tabular}
    }
    \end{threeparttable}
\end{table}

\section{The Ionization Properties and Structural Features of the Rosette Nebula}

\subsection{Spatial Distribution of the Optical Line Fluxes and their ratios}
\begin{figure*} 
    \centering
    \includegraphics[width=\textwidth]{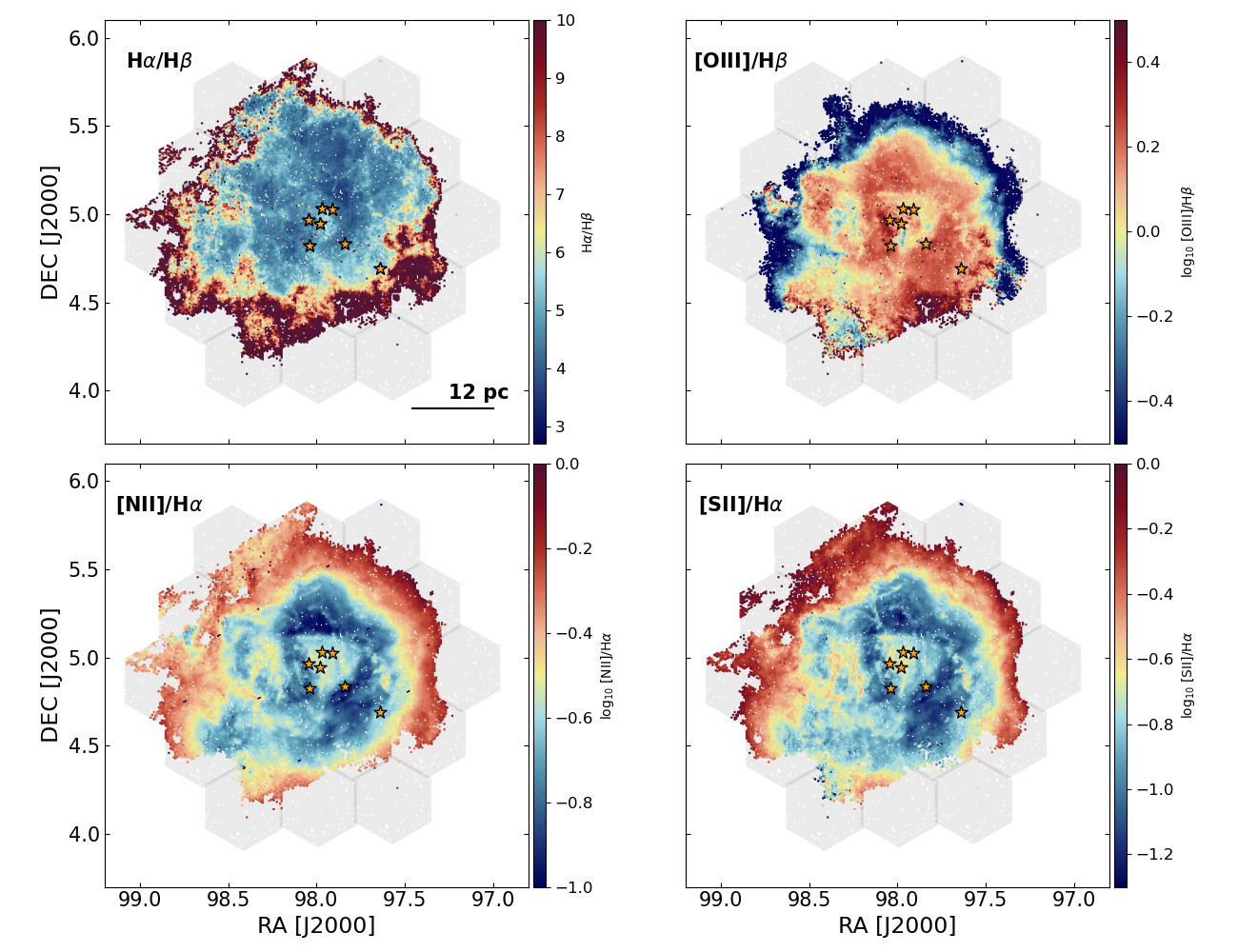} 
    \caption{Maps of the line ratios \Halpha/\Hbeta, \nii/\Halpha, \oiii/\Hbeta, and \sii/\Halpha\ in the Rosette Nebula. O-type stars are marked with yellow stars. Color bars indicate the relation in relative flux intensity. The scale bar in the \Halpha/\Hbeta\ panel represents 12 pc, assuming a distance of 1.5 kpc to the nebula. In these maps, only spaxels with a signal-to-noise ratio (S/N) $\geq$ 1 in the emission lines and S/N $\geq$ 10 in \Halpha\ were considered}
    \label{fig:elines}
\end{figure*}

Through the analysis of the spatial distribution of the different strong emission lines, we can analyze the physical properties of the nebula. However, due to biases in the current data calibrations mentioned earlier, these quantitative analyses are not included in this work. Nevertheless, the fluxes and the ratios between emission lines, such as \oiii/\Halpha\ and \sii/\Halpha, provide key information about the physical structure and ionization conditions in the nebula. Therefore, in this work, we focus on a qualitative description of the 2D distribution of the fluxes and their ratios to better understand the morphology of the region in light of previous studies.

Figure \ref{fig:Rosetta-RGB} presents the general distribution of the emission lines of \oiii, \Halpha, and \sii\ in relation to the \(12 \, \mu\text{m}\)  dust emission. The central cavity, in which the OB stars of the NGC 2244 cluster are located, is clearly observed. This cavity has been carved out by the radiation and stellar winds of the O-type stars belonging to the cluster, forming an expanding HII region \citep{Bruhweiler2010}.

The \oiii\ and \Halpha\ lines trace the  high-ionization region in the nebula, outlining an ionization ring with an approximate radius of 5 to 15 pc. Towards the edges of this structure, the \sii\ emission becomes more prominent, highlighting the mid-ionization regions and sharply delineating the transition between the ionized gas and the remnant molecular cloud. The \sii\ emission is particularly useful for identifying lower-ionization regions, as its ionization potential range is relatively low (10.36-23.40 eV) compared to \oiii\ (35.12-54.89 eV).

Additionally, the figure reveals various structures formed by the interaction between the expanding \Hii\ region and the surrounding molecular gas. Among these structures, filaments, protuberances ("elephant trunks"), and globules stand out. In particular, the structure known as the "wrench," as described  in previous studies \citep[e.g.,][]{Gahm2006, Makela2014}, is observed in the northeastern region. These structures are discussed in greater detail in the next sections. These morphological elements reflect the dynamics of stellar feedback and its impact on the evolution of the interstellar medium.

To describe the ionization structure of the Rosette Nebula, we analyze the spatially resolved emission-line ratios derived from the LVM data (Figure \ref{fig:elines}). The \Halpha/\Hbeta\ ratio map does not exhibit prominent morphological substructure, as this ratio primarily traces dust extinction along the line of sight through the Balmer decrement. The map shows generally low extinction values across the nebula, with a gradual increase toward the outer regions, likely associated with residual gas and dust from the remnant molecular cloud. 
A notable feature in the \Halpha/\Hbeta\ map is a clear north–south asymmetry in extinction, with higher values observed toward the southern part of the nebula. This asymmetry is consistent with the inclined geometry of the molecular ring proposed by \citealt{Dent2009}, who modeled the ring with an inclination of approximately $30^\circ$ with respect to the plane of the sky to reproduce the large-scale morphology structure of the molecular cloud. In this configuration, the southern side of the ring is tilted toward the observer, meaning that ionizing radiation from the central stars illuminates the ring from behind in the south, passing through more intervening dust, while in the north the radiation is seen emerging from the front side of the ring, resulting in lower observed extinction. This geometrical configuration naturally explains the extinction asymmetry observed in the \Halpha/\Hbeta\ map. In contrast to the \oiii/\Hbeta, \nii/\Halpha, and \sii/\Halpha\ maps, the O-type stars appear embedded within a region of uniformly low \Halpha/\Hbeta\ values, suggesting that these ionizing sources are located in areas of minimal foreground extinction.

In comparison with \Halpha/\Hbeta, the \oiii/\Hbeta\ ratio exhibits a significantly different distribution. This ratio shows a ring-like spatial morphology, with an inner radius smaller than that observed in the maps of the individual fluxes (see Appendix \ref{sec:appendix_A}). Additionally, it shows a relatively homogeneous value within the highly ionized region, corresponding to the interior of the nebula.
However, at the edges of the ionized region, the \oiii/\Hbeta\ ratio drops abruptly, precisely in the areas where the \sii/\Halpha\ and \nii/\Halpha\ ratios are higher. This may indicate a variation in the ionization degree at the boundary of the \Hii\ region. 

The \nii/\Halpha\ and \sii/\Halpha\ maps exhibit comparable morphological features, with small differences between the ratios. The \sii/\Halpha\ ratio is slightly lower in the inner regions of the nebula, while towards the edges, its value increases. This behavior may be related to the lower ionization potential of \sii\ with respect to \nii\, indicating a transition between the highly ionized gas region and the surrounding medium with lower degree of ionization. Both ratios increase with distance from the nebula's center, reaching their highest values at the boundary regions interacting with molecular gas.

Our analysis shows that most emission-line ratios trace comparable large-scale structures in the nebula. In Section \ref{sec:radial}, we further analyze the radial behavior of these line ratios and their correlation with other interstellar medium tracers, such as thermal dust emission.

\subsection{Morphological Comparison with Other ISM Phases and Stars}
\label{sec:comp}

\begin{figure*}[h!] 
    \centering
    \includegraphics[width=1\textwidth]{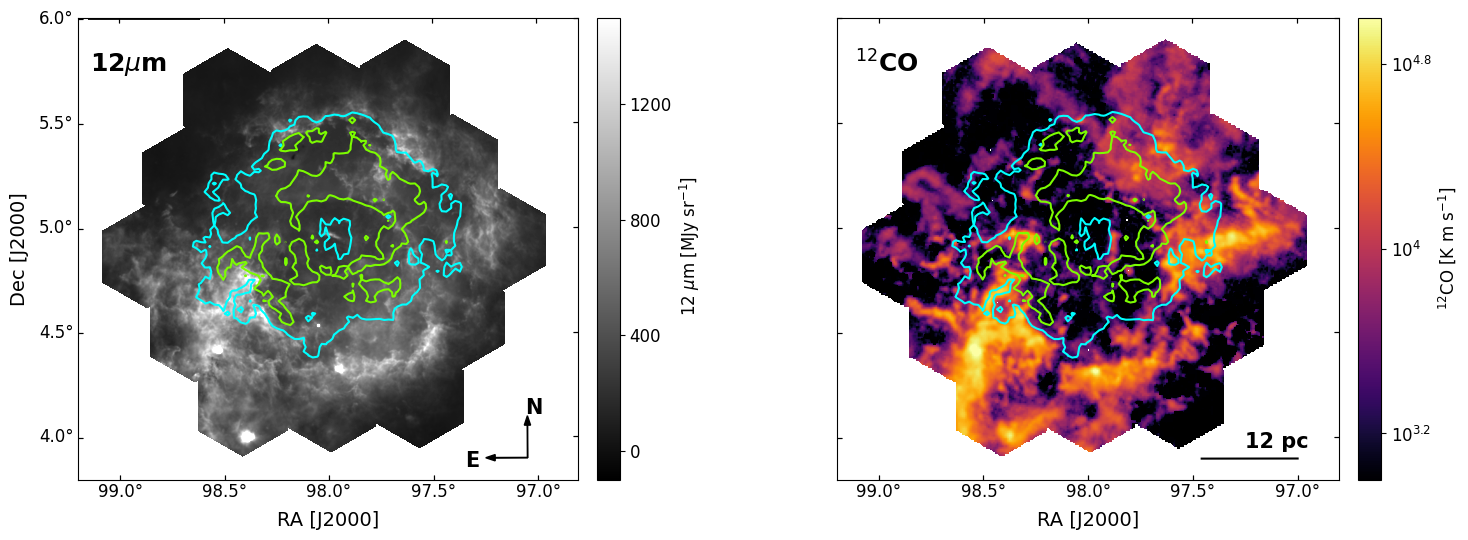} 
    \caption{Emission maps of the molecular cloud in \(^{12}\text{CO}\) (right) and dust in the \(12 \, \mu\text{m}\) band (left) in the Rosette Nebula region. In both images, the contours corresponding to 20\% (cyan) and 70\% (green) of the maximum relative \Halpha\ flux. The \Halpha\  contours highlight regions of ionized gas surrounding areas of molecular and dust emission, providing information on the interaction between the ionized gas and the gas and dust components in the molecular cloud. The intensity scale in each map is shown with color bars, indicating the emission in units of \( \text{MJy sr}^{-1} \) for dust (left) and \( \text{K km s}^{-1} \) for \(^{12}\text{CO}\) (right).}
    \label{fig:Mapas_ISM}
\end{figure*}

To better understand the interactions between massive stars and their local environment, it is essential to compare different tracers that allow us to analyze ionized gas, molecular gas, and dust. In this section, we focus on the morphology of the ionized region and how it relates to the structures observed in molecular gas and dust emission. We performed a morphological comparison of the relative flux obtained in \Halpha\ with the integrated intensity map of \(^{12}\text{CO}\) (J=1-0) (hereafter CO) and the thermal dust emission at \(12 \, \mu\text{m}\). In Fig.~\ref{fig:Mapas_ISM}, we superimpose the contours of the \Halpha\ emission on the maps of these two wavelengths.

The \Halpha\ contours delineate the structure of the ionized region, revealing an approximately concentric ring-shaped morphology with a characteristic size of $\approx$30 pc. At the center of this cavity are four O-type stars associated with the NGC 2244 cluster \citep{Lim2021}. The qualitative comparison with the CO and \(12 \, \mu\text{m}\) maps shows a clear contrast between the ionized region and the lower-intensity areas, which exhibit lower integrated intensity and weaker thermal emission, where the stars have evacuated the molecular gas. In particular, the dust defines the barrier where stellar radiation has not been able to fully penetrate the molecular gas, favoring the formation of structures such as globules or elephant trunks \citep{McLeod2015}.

The boundaries between  \Halpha\ emission and  \(12 \, \mu\text{m}\) dust emission generally coincide with the edges of the PDR, particularly in the southeastern quadrant of the nebula, where an increase in molecular gas density is observed. This behavior becomes even more evident when the integrated intensity of CO is compared to the location of the molecular cloud. In contrast, towards the northeast, both gas and dust densities are lower, which coincides with a decrease in the relative \Halpha\ flux, probably due to photoevaporation driven by the NGC 2244 cluster.

This pattern suggests that the structure of the primordial molecular cloud was quite inhomogeneous, allowing ionized gas to escape toward regions of lower density. Additionally, the observed differences in gas and dust emission support the hypothesis that the nebula formed within a filament of the original molecular cloud, extending in the southeast–northeast direction, where the presence of molecular material is more prominent. In a follow-up kinematic study we will provide a more detailed analysis of the gas dynamics in this region and its interaction with stellar radiation, providing a more comprehensive view of the nebula's evolution.

\begin{figure*} 
    \centering
    \includegraphics[width=1\textwidth]{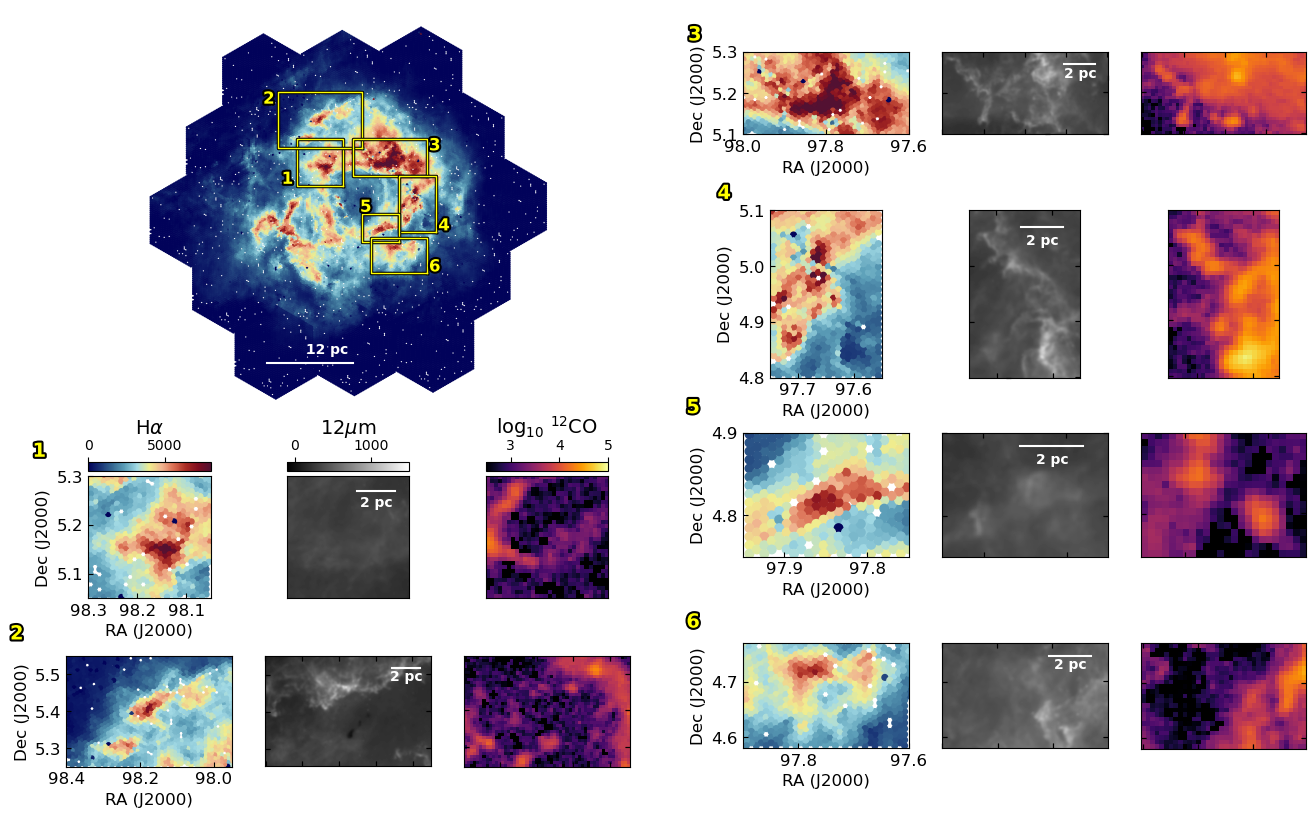} 
    \caption{Distribution of relative \Halpha flux, \(12 \, \mu\text{m}\)  dust, and CO in selected regions of the Rosette Nebula. The top left panel shows the selected areas in the \Halpha\ map, with six regions of interest marked with yellow rectangles. Each numbered region (1–6) is enlarged in the bottom and side panels, where \Halpha\ maps (left), \(12 \, \mu\text{m}\) dust emission (center), and CO emission (right) are shown. This selection of areas allows for a more detailed understanding of the spatial and morphological variations between the ionized and neutral gas components.}
    \label{fig:sel7}
\end{figure*}

To conduct a more in-depth analysis of the boundaries between molecular gas, dust, and ionized gas regions, we selected 6 areas that exhibit peaks in relative \Halpha\ flux (see Figure \ref{fig:sel7}). In this selection, we excluded the southeastern quadrant for a later analysis with Herschel data. These areas provide key information about the structure of the interstellar medium and the interaction of stars with it. They can be divided into two groups. The first group comprises areas 1, 2, and 5, which do not show a clear qualitative correlation or anticorrelation between highly ionized regions and those dominated by dust and molecular gas. The second group includes areas 3, 4, and 6, which exhibit a clear correlation between regions of higher molecular gas and dust density and areas of lower ionization.

Area 1 shows a well confined peak of \Halpha\ emission density, where the intensities of \(12 \, \mu\text{m}\) and CO are, in contrast, very low. This could indicate intense ionization activity in a region devoid of molecular gas. The dispersion and/or photoevaporation of the material, induced by the radiation from ionizing stars, could explain the scarce presence of CO and warm dust, but it also could be a region where the initial density of molecular gas was lower. This region is located within the ionization ring, which would justify the high \Halpha\ intensity due to its proximity to ionizing sources. In the same quadrant we highlighted Area 2, characterized by several smaller peaks of high \Halpha\ emission, accompanied by a discrete dust emission region towards the northern edge, where \Halpha\ emission decreases significantly. The CO emission is also low in this region. Located on the outer part of the ionization ring, the distribution of \(12 \, \mu\text{m}\) could reflect the effect of pressure exerted by the expanding nebula, which heats the dust at the periphery of the ionized gas. However, the scarce detection of CO may confirm that this quadrant had a low initial molecular gas density.

Area 5 is located on the inner edge of the ionization ring. Similar to Area 1, it exhibits medium-high intensity in \Halpha\ and medium-low intensity in CO and at dust. The low CO and \(12\,\mu\text{m}\) emission could be due to the rapid dispersal of material by the ionizing stars. However, this area shows a slightly higher presence of CO and dust compared to Area 1, which could be attributed to differences in the density of the progenitor molecular cloud and/or to weaker interaction with the stars of NGC 2244. The quadrant-based analysis in Section \ref{sec:radial} may provide further insights into the ionization processes occurring in this region and the observed emission intensities at different wavelengths.

On the other hand, in Areas 3, 4, and 6, the emission peaks for the different components of the interstellar medium do not coincide. The areas with low \Halpha\ emission exhibit higher thermal emission at \(12 \, \mu\text{m}\) and CO. The flux distribution between \(12 \, \mu\text{m}\) and CO is notably similar, tracing the remnant of the molecular cloud. These regions reveal the external structure of the ionized gas ring at the interface with the molecular cloud.

Area 3 contains a large fraction of the intense \Halpha\ emission in the northwestern quadrant of the map. It is also characterized by the presence of filamentary structures and peaks of high dust and molecular gas density (elephant trunks and globules.) These structures have been the subject of previous studies aimed at understanding their formation and evolution. For example, \cite{Makela2014} used near-infrared images, while \cite{Gahm2007} studied this region using images in optical bands. Both works identified that the globules are high-density regions and estimated their masses, finding that estimates based on infrared data are higher than those obtained from optical observations. In most cases, masses on the order of tens to hundreds of $\text{M}_\text{Jup}$ were found. Additionally, \cite{Makela2014} identified two young stellar objects (YSOs) in the region known as "the claw" ($\alpha = 97.9^\circ$, $\delta = 5.2^\circ$). The region called "the wrench" ($\alpha = 97.9^\circ$, $\delta = 5.12^\circ$) stands out on the \Halpha\ map as a thin absorption structure (spaxels in blue tones in our scale), in contrast to the extended surrounding emission (predominantly red-orange). These structures are clearly visible in the \(12 \, \mu\text{m}\) map, and CO observations confirm that it is also a region with a high molecular gas density.

Area 4 is of particular interest because it hosts the center of the NGC 2237 cluster. Previous studies have shown that this region is an active site of star formation. For example, \cite{Makela2017} identified young stellar objects (YSOs), which confirmed star formation activity in the region. We observe a high density in CO and dust emission in areas with lower \Halpha\ emission. This pattern suggests that the expansion of the ionized gas could be compressing the surrounding molecular gas, potentially inducing the formation of new stars. The NGC 2237 cluster is discussed in more detail in Section \ref{sec:ngc2237}.

Area 6 marks the boundary between ionized gas and a high-density region of molecular gas and dust. Its behavior is similar to that of region 4, showing a clear anticorrelation between \Halpha\ emission and \(12 \, \mu\text{m}\) and CO emission. This distribution highlights the interface between the ionized region and the molecular cloud. The pressure exerted by the expansion of ionized gas could be compressing the molecular gas, thereby favoring the formation of dense gas and dust structures.

\subsection{Morphological Comparison with the Column Density in the South East quadrant}
\label{sec:columndensity}

Using far-infrared maps from the Herschel Space Telescope, it is possible to construct a map that traces the column density within the molecular cloud \citep{Lombardi+14,Zari+16,HuYue21}. The comparison between the column density of molecular gas and ionization flux provides valuable information about the boundaries between the two phases of the ISM, highlighting the regions where high dust density shields the molecular cloud from photoevaporation. Based on the far-infrared maps of the Rosette Molecular Cloud carried out by the Herschel Space Observatory. These data have been used to construct column density maps, providing insights into the dynamics of the molecular cloud in relation to star formation \citep{Moran+2024}.

\begin{figure} 
    \centering
    \includegraphics[width=0.45\textwidth]{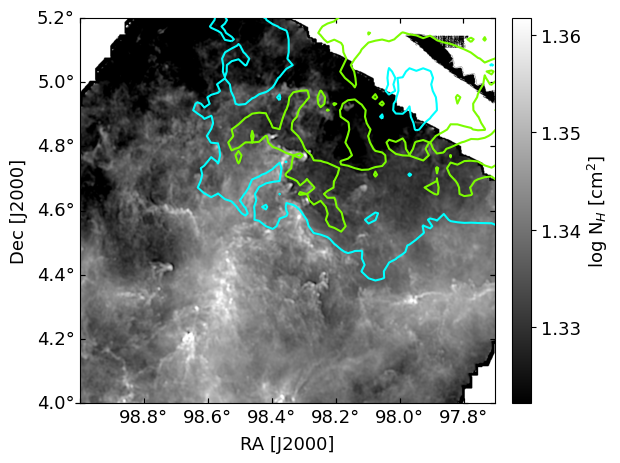} 
    \caption{Column density map derived from infrared data obtained with the Herschel Space Telescope, limited to the area overlapping the Rosette Nebula. \Halpha\ contours at 20\% (cyan) and 70\% (green) of the  maximum relative flux highlight the ionized gas distribution. These contours indicate the distribution of ionized gas in relation to areas of high column density, allowing for analysis of the interaction between the ionized regions and the denser molecular cloud structures. The color scale on the right displays the logarithmic values of hydrogen column density (\( \log N_{H} \)), measured in \(\text{cm}^{-2}\).}
    \label{fig:Herschel_Contour}
\end{figure}

In Figure \ref{fig:Herschel_Contour}, we present a combined map, which illustrates the interaction between the \Hii\ region traced by the LVM data (cyan and green contours) and the surrounding molecular cloud traced by the Herschel column density map. The image clearly reveals the boundary between the molecular cloud and the \Hii\ region: low-density areas are associated with ionized gas, while high-density areas correspond to less ionized regions. 
This suggests that the expansion of the \Hii\ region is ablating the molecular gas, thereby creating an environment conducive to the formation of globules and filaments, as radiation and stellar winds from OB stars ionize and/or expel the surrounding neutral material \citep[e.g.][]{McLeod2015, Mellema06}.

Furthermore, it is notable how the low-flux morphological structures, traced from the \Halpha\ emission (i.e., the cyan contours), are highly consistent with the structures with high column density. To illustrate this more clearly, In Figure \ref{fig:A_herschel} of the appendix, we present another inset comparison of \Halpha\ and column density, which highlights a low-density region with a relative ionization value of approximately $1\%$ compared to the maximum value of the LVM map. This underscores the high spatial resolution achieved with LVM, especially when compared to other high-resolution studies, such as those from Herschel.

\begin{figure*} 
    \centering
    \includegraphics[width=1\textwidth]{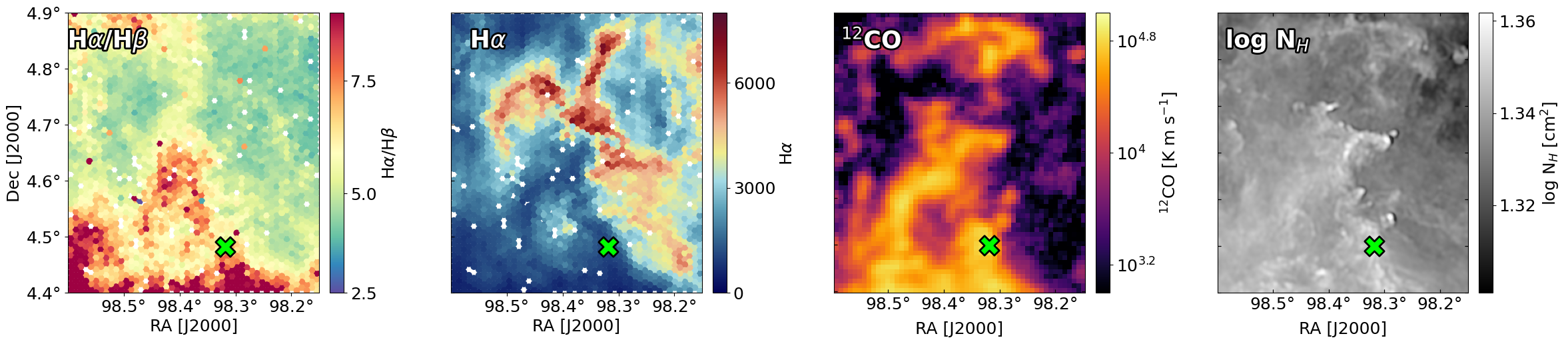} 
    \caption{Comparison of the \Halpha/\Hbeta\ ratio, \Halpha\ flux, molecular gas, and column density. The left panel displays the \Halpha/\Hbeta\ ratio map. The second panel shows the relative \Halpha\ flux. The third panel presents the \(^{12}\)CO emission map. The right panel shows the column density map in logarithmic scale (log \(N_{\mathrm{H}}\)). The green ''X'' indicates the position of the embedded cluster [PL97] 2. This comparison reveals that regions with enhanced \Halpha/\Hbeta\ ratios coincide with areas of higher column density, associated with dense CO clumps.}
    \label{fig:hahb}
\end{figure*}

In Figure \ref{fig:hahb}, we compare \Halpha/\Hbeta, \Halpha, the integrated CO intensity, and the column density derived from far-infrared data. This comparison focuses on the southeastern region of the nebula, where high CO intensity, elevated column density, and a high \Halpha/\Hbeta\ ratio are observed, in qualitative anticorrelation with \Halpha\ emission. This suggests a physical environment dominated by the interaction between ionized and molecular gas. 
The qualitative anticorrelation between extinction and \Halpha\  emission indicates that regions with higher dust and molecular gas density act as shielding boundaries that attenuates stellar radiation.
In the \Halpha\ map, an arc-shaped structure is observed at $\alpha = 98.2^\circ$, creating a local cavity characterized by low intensity across all tracers. Additionally, an \Halpha\  structure bordering this cavity is evident, suggesting a possible interaction between ionized and neutral gas with the stars of the NGC 2244 cluster. This interaction will be analyzed in Section \ref{sec:stars}.
High column density regions coincide with the morphological structures observed in CO, where areas of greater integrated CO intensity correspond to denser regions. Such regions  could represent gas clumps conducive to star formation, as the presence of dense molecular gas and the shielding effect of dust create favorable conditions for gravitational collapse and the onset of star formation.

Several studies have analyzed embedded star clusters located in high column density regions within the Rosette molecular cloud \cite[][hereafter PL97]{PL97}; \citep{Poulton08}. Other authors \citep[e.g.,][]{Roman2008,Hennemann10} have investigated clusters that are still undergoing active star formation processes. The evolution of the Rosette complex suggests that these clusters may have formed as part of the natural development of the molecular cloud \citep{Ybarra13}. However, there is also evidence that radiation from the PDR has influenced the formation of new stellar clusters \citep{Schneider10}.
An example of this process is the young embedded cluster [PL97] 2, marked with a green cross in Figure \ref{fig:hahb}. This cluster is located in a region characterized by higher CO intensity, elevated column density, and a high  \Halpha/\Hbeta\ ratio, along with relatively low \Halpha\ emission, precisely at the interface between the nebula and the molecular cloud. \cite{Roman2008} concluded that this cluster exhibits lower star formation activity compared to other embedded clusters in the region. This may be related to its proximity to the ionizing region, where intense radiation can ionize and ablate the surrounding material, thereby reducing the efficiency of new star formation.

Another example of an embedded cluster located within the interaction region between ionized and molecular gas is NGC 2237, situated in the western region of the nebula. This cluster is particularly interesting due to its position at the boundary between the ionized region and the molecular cloud, where external radiation may influence its star formation activity. A more detailed analysis of NGC 2237 is presented in Section \ref{sec:ngc2237}.

\subsection{Radial Distribution of Emission Line Ratios}
\label{sec:radial}

\begin{figure}
\centering
\begin{subfigure}[b]{0.8\linewidth}
\includegraphics[width=\linewidth]{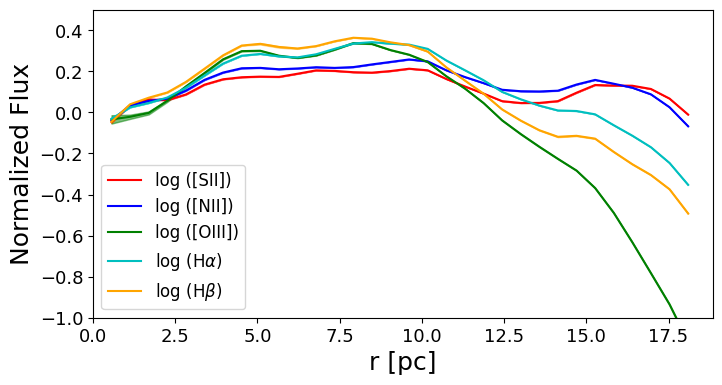}
\caption{}
\label{fig:Flux_elines}
\end{subfigure}
\begin{subfigure}[b]{0.8\linewidth}
\includegraphics[width=\linewidth]{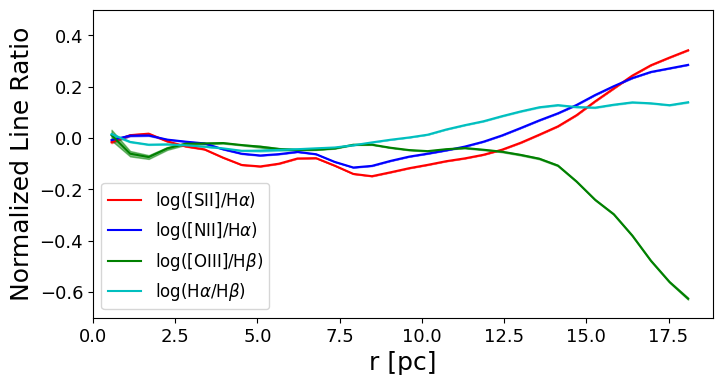}
\caption{}
\label{fig:Flux_relines}
\end{subfigure}
\caption{Radial variation of normalized emission line fluxes from the geometric center of the Rosette Nebula. (a) Normalized flux profiles of \sii, \nii, \oiii, \Halpha, and \Hbeta, showing an enhancement within the central 4.5 pc and a decline beyond 10 pc. (b) Normalized radial profiles of emission line ratios: \sii/\Halpha, \nii/\Halpha, and \oiii/\Hbeta, where the ratios with \sii\ and \nii\ highlight medium-ionization zones, while \oiii\ indicates the high-ionization region \citep{Kewley2019}}
\label{fig:Flux_radial_LVM}
\end{figure}

To better understand the spatial distribution of the gaseous component traced by the various emission lines, we decided to compare the normalized relative flux variations from the geometric center of the nebula, taking advantage of the apparently circular symmetry projected onto the plane of the sky. As mentioned in \cite{Bruhweiler2010}, this spherical geometry, as a first approximation, allows for the analysis of RN under the assumption of a nearly uniform expansion of the \Hii\ region. This expansion is driven by the combination of stellar ionization, the dynamics of the expanding bubble, and the interaction with the surrounding gas, which governs the structural and kinematic evolution of the interstellar medium within the nebula. We define the coordinates $\alpha = 98.0^\circ$ and $\delta = 4.9^\circ$ as a geometric center and, we calculated the mean normalized emission in concentric rings with a separation of 0.72 pc and a width of 1.44 pc. These values were chosen so that each ring would cover an area equivalent to approximately three fibers in width, ensuring both spatial continuity and statistical robustness in the resulting profiles. The normalized flux values were obtained by dividing them by the central flux at the geometrical center for each tracer\footnote{We estimate the central flux using the values of different tracers within the coordinate range $\alpha = 97.7^\circ$ - $98.03^\circ $ and  $\delta = 4.87^\circ$ -$ 4.93^\circ$.}, allowing for an analysis of radial trends in the different emission lines and a morphological comparison across the various wavelengths. The resultant profiles are shown in Figure \ref{fig:Flux_radial_LVM}. 

The normalized fluxes of the emission lines of \sii, \nii, \oiii, \Halpha, and \Hbeta\ are shown in Figure \ref{fig:Flux_elines}. A significant increase is observed from the center out to $\sim$ 4.5 pc, corresponding to the edge of the central cavity associated with the stellar bubble formed by the ionizing stars of NGC 2244. This region marks the dense and highly ionized core of the nebula. Beyond this, a gradual decline in flux is observed, particularly beyond $\sim$ 10 pc, which likely marks the boundary of the most strongly ionized region. \citet{Bruhweiler2010} estimated a size of approximately 6 pc for the central cavity. The emission lines showing the most significant flux increase are \oiii, \Hbeta, and \Halpha, which follow a similar trend, whereas \sii\ and \nii\ show a more modest increase.
At larger radii, the fluxes of the \oiii, \Hbeta, and \Halpha\ lines decrease, while \sii\ and \nii\ show a slight increase around $\sim$ 14 pc. The \oiii\ and \Halpha\ lines are present in high-ionization regions, considering that the ionization potential of \oiii\ is 54.83 eV. In contrast, \nii\ and \sii\ correspond to medium-ionization lines, with significantly lower ionization potentials.

The flux ratios are shown in Figure \ref{fig:Flux_relines}. The graph shows that the behaviour of all the ratios is approximately uniform up to about 10 pc,  where the \sii/\Halpha\ and \nii/\Halpha\ ratios depart to higher values with a similar trend, while \oiii/\Hbeta\ exhibits an opposite behaviour, with a dramatic decrease.The trends are consistent with our interpretation of the emission lines discussed above, highlighting a central region of high ionization and an outer zone dominated by intermediate-ionization conditions \citep{Kewley2019}.

\subsection{Angular Distribution of Interstellar Medium Tracers}

\begin{figure} 
    \centering
    \includegraphics[width=0.45\textwidth]{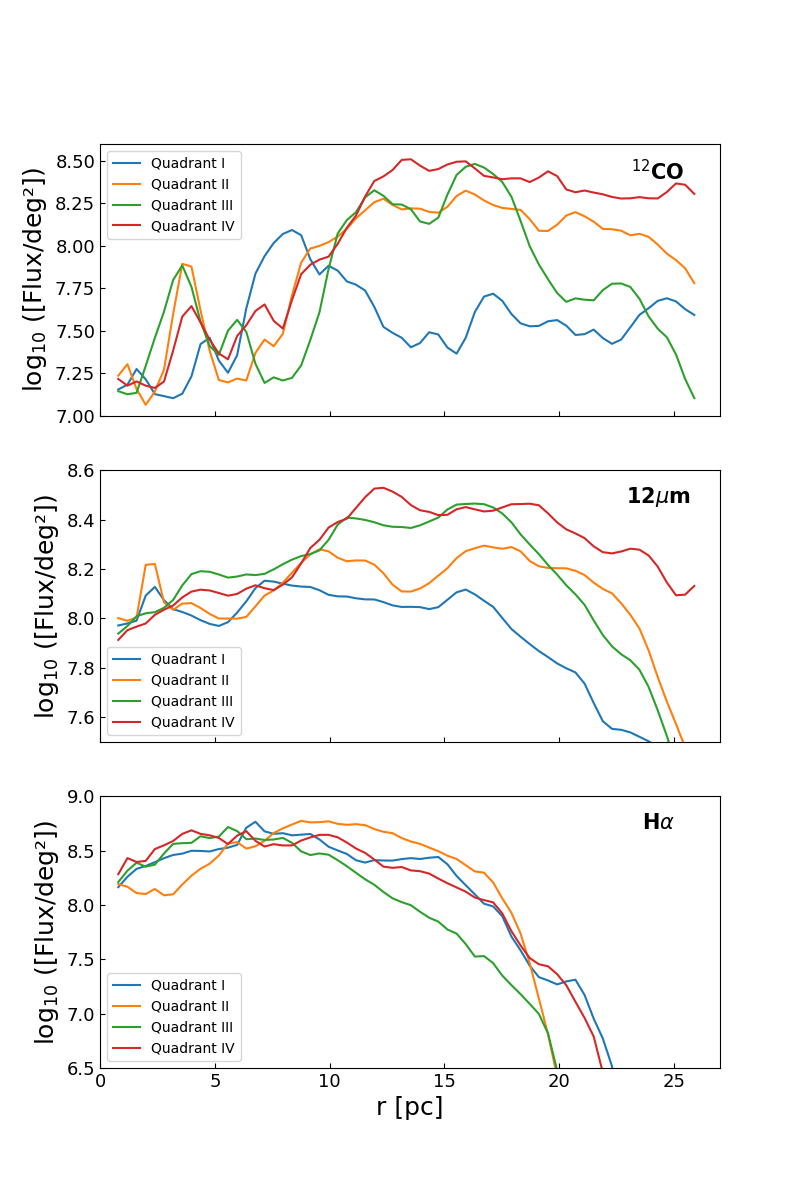} 
    \caption{ Radial distribution of emissions in the Rosette Nebula for different quadrants. The panels show the radial flux profiles for the emissions of \(^{12}\)CO (top), \(12 \, \mu\text{m}\) dust (middle), and relative \Halpha\ flux (bottom).  Colors represent the different quadrants: I (blue), II (orange), III (green), and IV (red), corresponding to the northeast, northwest, southwest, and southeast regions of the nebula, respectively. The radial distance r is measured in parsecs from the center of the nebula ($\alpha = 98.0^\circ ,\ \delta = 4.9^\circ$). This comparison reveals how the distribution of molecular gas, dust, and ionized gas varies with direction and distance from the center of the nebula.}
    \label{fig:Flux_CO_Dust_LVM}
\end{figure}

To gain a more detailed understanding of the spatial distribution of various interstellar medium (ISM) tracers, $^{12}$CO,  \(12 \, \mu\text{m}\) and relative \Halpha\ flux , we computed the flux per unit area in concentric radial rings, following the methodology described in the previous section.  To explore the spatial variations in more detail, each map was then divided into four quadrants by two imaginary lines intersecting at the geometric center. The quadrants are defined as follows: the first corresponds to the northeast, the second to the northwest, the third to the southwest, and the fourth to the southeast. This approach allowed us to analyze the flux evolution for different medium components as a function of radius in each region. The resulting radial distribution can be seen in Figure \ref{fig:Flux_CO_Dust_LVM}.

In the first quadrant, the integrated intensity of CO and  \(12 \, \mu\text{m}\) emission are lower compared to the other quadrants. This region could represent lower-density areas in the progenitor molecular cloud, potentially indicating a window through which ionized gas has escaped to the exterior of the nebula. This suggests the presence of non-uniform structures within the nebula.
In the second and third quadrants, the presence of CO is detected within the cavity. This small but significant integrated intensity of CO could represent gas expanding perpendicularly to the plane of the nebula, located either in front of or behind it. Alternatively, it could be gas that has not yet been completely dispersed. A detailed kinematic analysis would help clarify this interpretation. Additionally, in the second and fourth quadrants, the integrated CO intensity becomes more uniform beyond 12 pc, with higher values in the fourth quadrant. This delineates the regions of higher molecular gas density, contrasting with the third quadrant, which supports the idea of a possible filamentary structure of the progenitor molecular cloud extending from the southeast to the northwest.

Upon reaching approximately 10 pc, a decrease in relative \Halpha\  emission is observed across all quadrants, more notably in the third and fourth quadrants, while the CO flux increases. This marks the regions of interaction between the intense radiation from ionizing stars and the molecular cloud. Beyond 15 pc, the relative flux \Halpha\ decreases rapidly, indicating the transition to the denser region of the molecular cloud, where ionizing radiation can not penetrate or ionize the gas. This clearly defines the boundary between the ionized gas and the molecular gas.
In the fourth quadrant, the remnant of the molecular cloud is more evident, with elevated CO and dust fluxes that increase beyond 10 pc and remain relatively constant. This indicates a greater presence of molecular material in this direction.

Across all quadrants, the dust profile is more uniform and exhibits less significant variations, decreasing at larger radii, with a less pronounced decline in the fourth quadrant.
In the first and second quadrants, the \Halpha\ profiles show smoother variations between 5 and 15 pc, followed by a significant decrease beyond 15 pc. This highlights the ring-like structure of the ionized gas. 
The CO profiles reveal the filamentary structure of the progenitor molecular cloud, extending from the fourth quadrant toward the second. Furthermore, the first quadrant appears to be the least dense, showing lower emission.

The profiles observed in the different quadrants in both \Halpha\ and CO suggest that the nebula is neither symmetric nor homogeneous. This appearance may be due to the presence of relatively scarce intervening material along the line of sight, allowing the ring-like structure projected onto the plane of the sky to be seen more clearly. A plausible explanation is that the Rosette Nebula lies close to the plane of the sky, with an inclination of approximately $\sim 30^\circ$, and that it formed within a thin molecular shell, as proposed by the models of \citet{Wareing2018}.
These variations indicate that the density in the region is far from uniform, adding complexity to the structural analysis. These results emphasize the importance of conducting a detailed kinematic study to better understand the morphology, dynamics, and expansion of the nebula.

\subsection{Diagnostic Diagrams of the Ionized Gas}

\begin{figure*} 
    \centering
    \includegraphics[width=\textwidth]{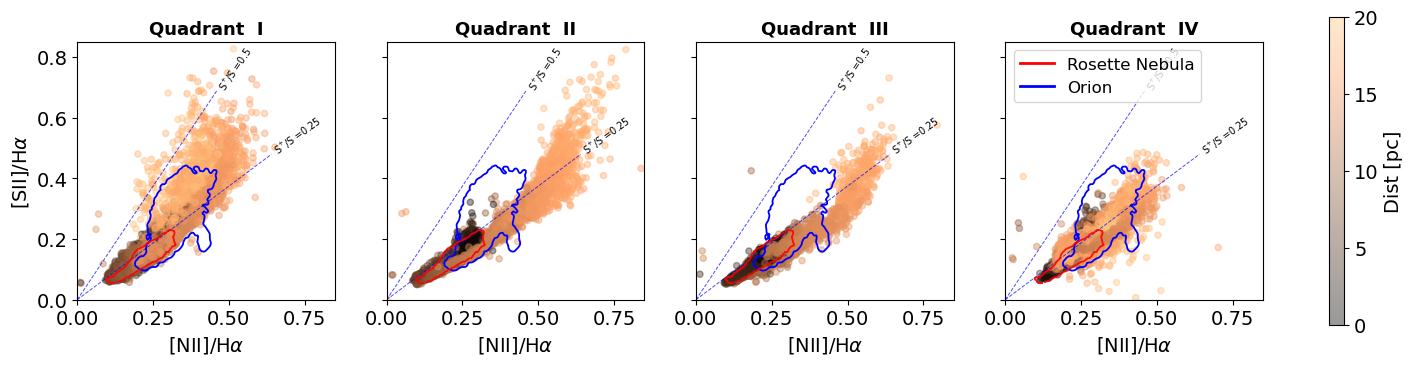} 
    \caption{ Ratio diagrams, \sii/\Halpha\ vs. \nii/\Halpha, for the four quadrants of the Rosette Nebula. The color scheme represents the distance from the center of the cavity in parsecs (pc). The blue dashed lines show theoretical values for different S\textsuperscript{+}/S ratios. The contours highlight the regions with the highest data density for Orion (blue) and the RN (red).}
    \label{fig:nhsh_c}
\end{figure*}

\begin{figure*}
\centering
\includegraphics[width=\linewidth]{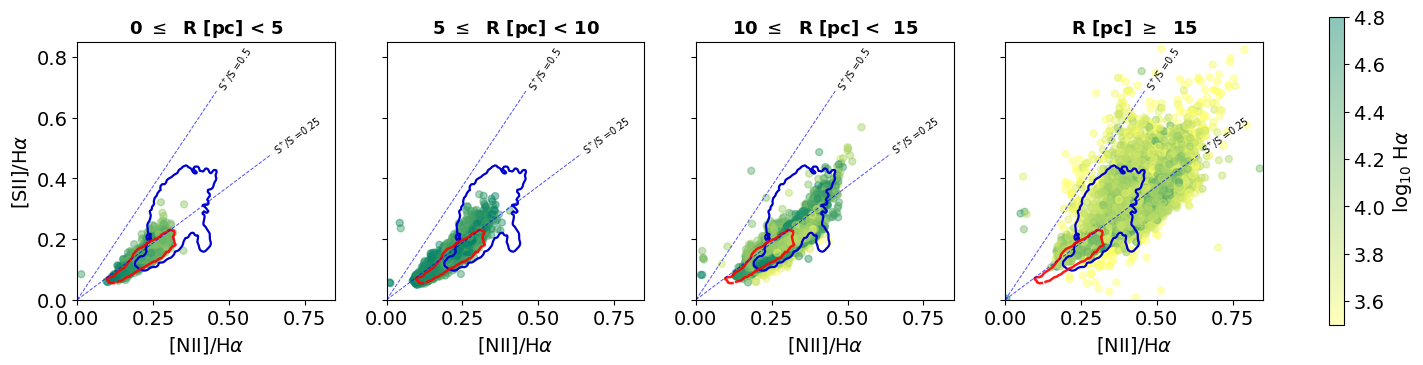}
\caption{Ratio diagrams for different radial distance ranges from the center of the cavity in the nebula. The color code represents the intensity of the relative \Halpha\ emission. The lines and contours are the same as those shown in Figure \ref{fig:nhsh_c}. We remind the reader that the LVM data presented in this study are still in a preliminary stage of scientific calibration. Therefore, our analysis is comparative, based on relative fluxes. A more detailed estimation of physical properties (e.g., T$_e$, n$_e$) will be presented in a follow-up study.}
\label{fig:nhsh_r}
\end{figure*}

The ratios between different emission lines provide a general understanding of the physical properties of the ionized medium. Diagrams such as \sii/\Halpha\ vs. \nii/\Halpha, or the classical BPT diagram, allow for an initial interpretation of the physical state of the ionized gas, for example, distinguishing between shocks or hot low-mass evolved stars, HOLMES, \citep{Sanchez2020}. These spectral diagnostics offer a first-order approximation of the different ionization mechanisms. Although traditionally applied to integrated spectra, LVM enables the acquisition of spatially resolved integrated spectra, offering a more comprehensive view of the physics and properties of HII regions.

The emission lines of \sii\ and \nii\ have ionization potentials that do not differ significantly, which makes the comparison between these lines largely independent of temperature \citep{Madsen2006}.  In Figure \ref{fig:elines}, we show how the distributions of \sii/\Halpha\ and \nii/\Halpha\ exhibit similar trends in both the radial and quadrant-based analysis. The radial profiles confirm that both ratios evolve similarly with increasing distance from the center of the nebula.
It is important to note that by analyzing the \sii/\Halpha\ values in relation to \nii/\Halpha\, we aim to obtain a qualitative and general understanding of the properties of the nebula. We are not performing a detailed calculation of physical parameters, as this would require comparisons with photoionization models and abundance determinations, which are beyond the scope of this study, as previously mentioned.

To interpret these results, in Figs. \ref{fig:nhsh_c} and \ref{fig:nhsh_r} we plot the \sii/\Halpha\ ratio as a function of \nii/\Halpha. The diagrams follow the approach of  \cite{Madsen2006}, applied to the warm ionized medium (WIM). Additionally, we followed the methodology of \cite{Kreckel2024} for the Orion Nebula, based on Equations 2  from \citeauthor{Madsen2006}, varying the value of N\textsuperscript{+}/N. 

In both studies, a value of N\textsuperscript{+}/N = 0.8 is adopted for the WIM, where the ionization is low. However, in an HII region, ionization is harder, and a significant fraction of nitrogen is expected to be in the N\textsuperscript{++} state, reducing the N\textsuperscript{+}/N ratio. Therefore, in our analysis, we use N\textsuperscript{+}/N = 0.4, which is more representative of the interior of the Rosette Nebula (RN). This fraction may vary throughout the RN, reaching values closer to those of the WIM in the outer parts of the nebula. Nevertheless, since this is a qualitative analysis, it serves as a first-order approximation. In each of the diagrams, the diagonal lines represent values of S\textsuperscript{+}/S obtained using Equation 2 from \citeauthor{Madsen2006}.

To analyze the structure of the RN, we employ two complementary approaches: a quadrant-based analysis, using the same quadrants as the radial profiles, which helps us understand the distribution and symmetry of the nebula, and a radial analysis, where we examine different distance intervals from the center of the nebula to identify variations in gas ionization and temperature.  

In all the diagrams, we show a contour enclosing the LVM data from Orion (in blue), obtained by \cite{Kreckel2024}, and the corresponding contour for the RN (in red). Both contours were constructed from a two-dimensional histogram highlighting the regions with the highest data density, considering 20$\%$ of the maximum histogram value along the x and y axes for both Orion and the RN. The physical resolution, and hence data density, in Orion is considerably higher, which results in greater dispersion and a larger area covered in the diagram.
The positions of both regions in the diagrams show a similar trend, although the RN exhibits lower line ratios. According to the literature, electron temperatures for Orion range from 7000 to 9000 K \citep{Wilson2015}, while lower values have been obtained for the RN, with estimates around 5800 K \citep{Celnik1985}. However, broader temperature ranges have also been reported, spanning from 3000 to 9000 K \citep{ReviewRoman2008}. The studied region in Orion shows weak \oiii\ emission, making it a more suitable representation of WIM conditions \cite{Kreckel2024}. According to the analysis by Madsen, WIM regions are expected to exhibit higher electron temperatures.

In Figure \ref{fig:nhsh_c}, shows the relationship between \nii\ and \sii, where colors represent the distance in parsecs from the nebula’s center for each spaxel. Each panel corresponds to a quadrant of the nebula, as described in section 3.5.

The first quadrant presents a dispersion similar to that observed in Orion and a lower density of points in the central region, indicating greater data dispersion. At larger distances, the \sii/\Halpha\ ratio is higher than \nii/\Halpha\, with a tendency toward higher S\textsuperscript{+}/S values. These variations could be primarily due to changes in the ionization degree or to variations in the gas density.

The second quadrant exhibits higher \nii/\Halpha\ values compared to the other quadrants. 
Furthermore, the data exhibit a linear trend up to a \nii/\Halpha\ ratio of approximately 0.6, beyond which a systematic increase in the  \sii/\Halpha\ ratio is observed for spatial positions located in the outer regions of the \Hii\ region, while remaining within the same \nii/\Halpha\ range.
This may be associated with the interaction between the ionized and molecular gas, suggesting the presence of shock effects or recombination. The third quadrant follows a similar trend, although less pronounced.

The fourth quadrant exhibits a greater dispersion of points within a more limited range of \nii/\Halpha\ values. At larger distances, the \nii/\Halpha\ and \sii/\Halpha\ values do not tend to increase. This quadrant is located in a region with greater interaction between the molecular cloud and the ionized gas, which slows down the expansion of the \Hii\ region. 

The Figure \ref{fig:nhsh_r} shows the distribution of \sii/\Halpha\ vs. \nii/\Halpha\ at different distance intervals from the center. In the innermost regions (0-5, 5-10 pc), a higher intensity of \Halpha\ is observed, leading to lower \sii/\Halpha\ and \nii/\Halpha\ values. These intervals exhibit a more linear trend and lower data dispersion, suggesting that the central part of the nebula has more homogeneous ionization compared to greater radii.
At increasing radii, greater data dispersion is observed, along with regions of lower \Halpha\ intensity and higher \sii/\Halpha\ and \nii/\Halpha\ ratios. The 10-15 pc interval shows a mixed structure, consistent with the fact that this region still maintains significant \Halpha\ intensity while also marking the interface limit with the molecular cloud. This further suggests that the region is not entirely homogeneous.
At larger radii ($>$15 pc), the highest dispersion of points is recorded, and \Halpha\ intensity decreases, possibly indicating a greater contribution from recombination processes. The radial distributions suggest that electron temperature is lower at smaller distances, possibly due to the efficient cooling provided by elements such as oxygen. In contrast, at larger distances, ionizing radiation decreases, favoring higher electron temperatures and enhancing the presence of collisionally excited emission lines.

\section{Interaction of young stars with their local environment}
\label{sec:discusion}

The LVM data, complemented by studies at other wavelengths, allow us to analyze the morphology and ionization structures produced by the interaction of OB stars with the molecular cloud. In an effort to better understand such structures, we selected several regions of interest. This allows us to comment on the evolution of embedded clusters and, in general, on the star formation processes. Although the entire complex serves as a laboratory for analyzing stellar feedback and its impact on star formation efficiency in molecular clouds, we focus on three specific areas.

\subsection{NGC 2237}
\label{sec:ngc2237}

NGC 2237 is a low mass young star cluster located in the western region of the Rosette Nebula. It is immersed in a complex structure of ionized gas, molecular gas, and dust. This cluster has been studied to understand the star formation processes in the region and the influence of high-mass stars as possible triggers. NGC 2237 lacks massive stars that could promote a rapid dispersal of the remnant molecular parental gas, and for that reason, the cluster is still deeply embedded, which is why it has been predominantly observed in the near-infrared \citep{Roman2008} and X-rays \citep{Wang2010}. The latter reference remains the most comprehensive study to date. In that study, 168 candidate cluster members were identified, of which seven belong to Class II and the rest to Class III, confirming that it is a young cluster still in the process of formation.

\begin{figure*}
    \centering
    \includegraphics[width=0.9\textwidth]{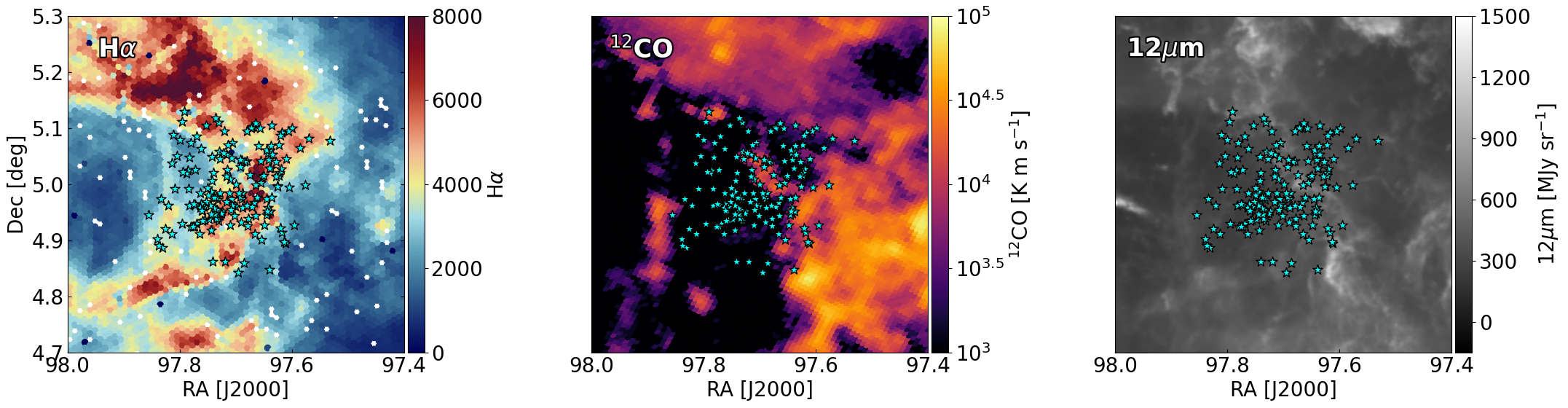} 
    \caption{Maps of the Rosette Nebula region around NGC 2237, showing the dust distribution at \(12 \, \mu\text{m}\) (left), \Halpha\ emission (center), and \(^{12}\)CO emission (right). The cyan points mark the positions of the 168 stellar members of the NGC 2237 association identified by \citet{Wang2010}, located in a region with complex structures of ionized gas, molecular gas, and dust.}
    \label{fig:ngc2237}
\end{figure*}

The LVM provides insightful data for studying the evolution of NGC 2237 and the influence that NGC 2244 has had on it. Figure \ref{fig:ngc2237} shows the spatial distribution of the cluster members \citep{Wang2010}, compared with the maps of \Halpha, CO, and dust It is important to note that the apparently rectangular geometry of the cluster in the figure is not intrinsic, but rather a consequence of the observational limitations of the study by \citeauthor{Wang2010}{}, which was restricted to a 17'$\times$17' field of view.  A large percentage of the members are located in a highly ionized region devoid of dust and gas. In the region centered at $\alpha = 97.68^\circ$ y $\delta = 5^\circ$, the stars are found to border the remnant of the molecular cloud, with a Class II star located at the tip of this remnant (towards the northeast). 
This region, known as the "Seahorse", was studied by \citet{Makela2017}, who identified young stellar objects (YSOs) of Class I/II. The region contains molecular gas and dust and is lightly ionized, suggesting that the dust acts as a barrier, preventing the ultraviolet radiation from NGC 2244 from fully ionizing the medium.

The true distance to NGC 2237 
has been debated in recent studies. Earlier works, such as \citet{Wang2010}, adopted a heliocentric distance of 1.4 kpc, under the assumption that NGC 2237 lies at the same distance as the central cluster NGC 2244. This choice is based on the estimated age of the cluster, about 2 Myr. However, more recent studies based on Gaia EDR3 data suggest that NGC 2237 may be located slightly farther away. In particular, \citet{Muzic2022} report a distance of 1440 $\pm$32 pc for NGC 2244 and 1525$\pm$36 pc for NGC 2237, implying a separation of approximately 85 pc between the two clusters along the line of sight. This result suggests that NGC 2237 is not strictly coplanar with NGC 2244, and that different regions of the Rosette complex may lie at varying depths
Assuming that the ionized gas and NGC 2237 are physically associated, most of the cluster members are observed to lie within already ionized regions. This implies that their formation likely occurred before the ionization from the OB stars in NGC 2244 expanded into this area. Although pressure from expansion of the H II region could have triggered star formation, some studies suggest that the age of the cluster does not differ significantly from that of NGC 2244 \cite{Muzic2022}. This supports the hypothesis that NGC 2237 formed contemporaneously with NGC 2244, rather than through a triggered star formation process.

\subsection{Intermediate mass stars}
\label{sec:stars}

\begin{figure} 
    \centering
    \includegraphics[width=0.4\textwidth]{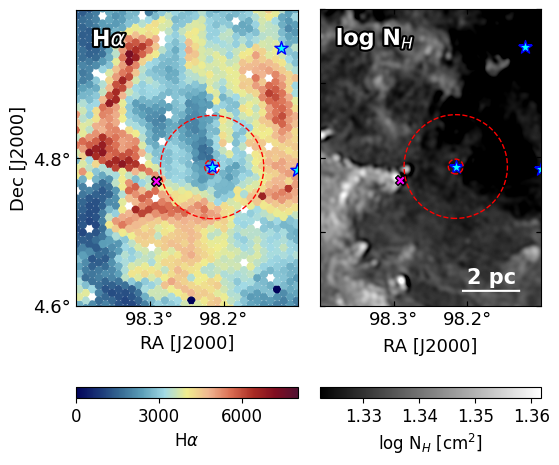} 
    \caption{Map showing the star NGC 2244 334 and its surrounding environment in the Rosette Nebula. The left panel displays the \Halpha\ emission, while the right panel shows the hydrogen column density distribution. A red dashed circle outlines an ionized gas shell structure, approximately 2 pc in diameter, with an arc-shaped feature oriented toward the center of NGC 2244. Cyan and blue stars mark the positions of nearby OB stars, highlighting the influence of the broader stellar association on this region. The magenta 'X' indicates the tip of a nearby globule}
    \label{fig:Herschel_Star}
\end{figure}

In Figure \ref{fig:Herschel_Star}, the star NGC 2244-334 can be observed, located at a distance of $1.50\pm0.05$kpc. This star appears to be surrounded by an ionized gas shell structure, with the western region resembling an arc directed toward the center of NGC 2244. The cavity has an approximate width of 2 pc. This region corresponds to an area of lower column density and a denser globule located east of the star, at approximately 1.9 pc. The star has been reported by \citet{Shang2022} as a B5V star, with a surface temperature of approximately $1.5\pm10^4$K. This temperature theoretically corresponds to a low rate of ionizing photons, with $N_{\gamma} = 10^{43}$. If we assume the gas is at a temperature of $10^4$K and the hydrogen density is $n_{H}=100 cm^{-3}$, we obtain a Str\"omgren radius of 0.015 pc. However, \cite{Bagnulo2004} report a temperature of approximately $1.6\pm10^4$K and classify the star as B3. Using the same gas parameters and $N_{\gamma} = 10^{45}$, the resulting Str\"omgren radius would be $Rs=0.147 pc$ . These values suggest that the observed structure could not have been formed solely by the influence of this star. To explain the observed ionization radius, a star of spectral type O8, with an effective temperature of $T_{eff} = 3.5 \pm 10^4$K, would be required. Therefore, this structure could be the result of the combined radiation from the OB stars in NGC 2244, or this star may have a higher temperature than previously estimated. 

Additionally, the arc-like structure may not be located at the same distance as the globule and could instead represent a misinterpretation of the molecular cloud structure due to projection effects.

\begin{figure} 
    \centering
    \includegraphics[width=0.4\textwidth]{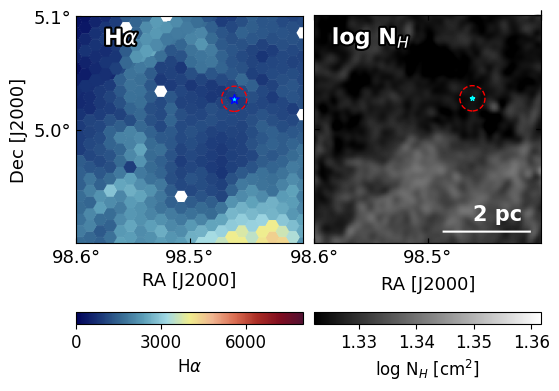} 
    \caption{Map showing the position of the eclipsing binary star LS VI +05 12 (spectral type B2.5V) in \Halpha (left) and column density (right), with its ionization radius (Rs=0.29 pc) marked. This star is located in a region to the east of the nebula, characterized by low ionization and column density, with a break in the molecular cloud.}
    \label{fig:Herschel_Binary_Star}
\end{figure}

We also highlight the region containing the eclipsing binary star LS VI +05 12, with a spectral type B2.5V, located at a distance of $1541.8\pm51.8$ pc. The orbital period of this star is approximately 25 days \citep{Pribulla2010}. In Figure \ref{fig:Herschel_Binary_Star}, the position of the star is shown in the \Halpha\ and column density maps, located to the east of the nebula. The circle indicates its ionization radius, Rs=0.29 pc.

This star is located in a region characterized by low ionization, surrounded by material with low to moderate ionization levels. This region corresponds to the northeastern quadrant of the nebula, identified as an area of low ionization and low density across different wavelengths (see Sections \ref{sec:comp}, \ref{sec:radial}). Consistently, the column density in this region is also low, although some denser structures are identified in the vicinity of the star.
The observed low densities could be related to a possible disruption or evacuation of material, suggesting that part of the gas has been expelled or ablated, possibly due to the action of stellar winds. \citet{Wareing2018} analyzed the morphology of the nebula using various models and concluded that the stellar winds from HD 46150 can induce bubble disruption in regions of lower density. This hypothesis requires a more detailed analysis, as well as a comparison with magnetohydrodynamic models, to assess whether the binary system in this area actively contributes to such disruption.
On the other hand, although no particularly high ionization is observed in this region, a high \oiii/\Hbeta\ ratio is detected. Understanding the origin of these values requires a more in-depth study using chemical abundance models, which lies beyond the scope of this work. Nevertheless, these results highlight the value of the data provided by LVM in characterizing the evolution and physical properties of \Hii\ regions, in complement to observations at other wavelengths.

\begin{figure} 
    \centering
    \includegraphics[width=0.5\textwidth]{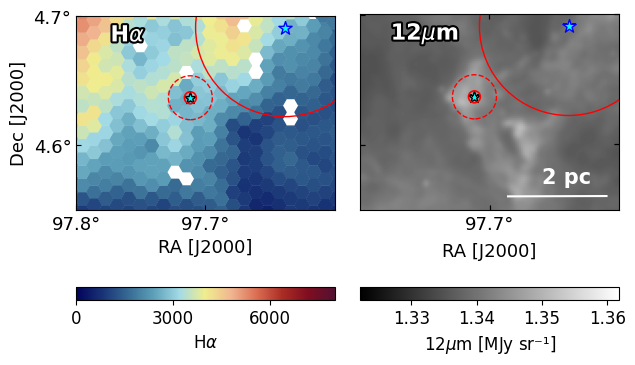} 
    \caption{Map showing the positions of the stars Cl NGC 2244 CDZ 36 (B5) and HD 258691 (O9.5V) overlaid on \Halpha\ (left) and dust emission (right) maps. Solid red circles indicate their respective ionization radii (Rs=0.12 pc and Rs=1.80 pc). The B5 star, located 2.36 pc southeast of HD 258691, is enclosed by a dashed red circle with a radius of 0.45 pc, corresponding to a region of low \Halpha\ emission and enhanced dust concentration. An arc-shaped dust structure is seen surrounding the B5 star, extending eastward.}
    \label{fig:Stars_OB}
\end{figure}

The star Cl NGC 2244 CDZ 36, with a spectral type B5, is located in the southeastern region of the nebula, at a projected distance of 2.36 pc from the O9.5V star HD 258691. Figure \ref{fig:Stars_OB} shows the positions of both stars over the \Halpha\ and dust maps, along with their estimated ionization radii, Rs = 0.12 and Rs = 1.80 pc, respectively.
A circle with a radius of 0.45 pc is drawn around Cl NGC 2244 CDZ 36, delimiting a region characterized by a combination of low \Halpha\ emission and high dust intensity. This suggests that the star is located in an area where the surrounding molecular cloud has not yet been fully ionized or ablated by stellar radiation. Additionally, an arc-shaped dust structure is observed to the east of the star, possibly indicating a partial clearing of material or the presence of a small bubble.
Although HD 258691 appears close in the plane of the sky, distance estimates and their corresponding uncertainties provided by Gaia DR3 (see Tables 1 and 2) indicate that Cl NGC 2244 CDZ 36 is likely located several hundred parsecs away, making a physical interaction between the two stars unlikely.
One possible interpretation is that the observed arc-shaped structure corresponds to a weak bow shock, formed by the interaction between the stellar wind of Cl NGC 2244 CDZ 36 and the large-scale wind from the massive central stars of NGC 2244, primarily HD 46150. This hypothesis remains speculative and should be further explored in future studies.

The selection of these regions shows that the ionization in the Rosette Nebula is primarily dominated by the most massive stars of NGC 2244.
According to \citet{Wareing2018}, the dynamics and morphology of the nebular bubble are controlled mainly by HD 46150 (spectral type O5V) 
However, in a more detailed analysis, the contribution of OB stars, which help disperse and shape the gas structures we observe in the RN, should not be overlooked.

A comparison of the kinematics of the stars and the various phases of the local ISM can provide a better understanding of the three-dimensional structure of the gas and the star, shedding light on the dominant physical processes in the region. This kinematic study will be addressed in a forthcoming paper (in preparation).

\section{Summary and Conclusions}

We conducted a morphological analysis of the Rosette Nebula at different wavelengths, emphasizing its ionization structure and the interaction between ionized gas and the surrounding molecular cloud. Using optical emission lines maps obtained from the LVM , we investigated the spatial distribution of ionized gas and compared it with thermal dust emission maps and integrated molecular gas intensity to better understand the processes governing the evolution of the region.
Comparative analysis between the structures traced by relative \Halpha\ emission and the dust and molecular gas maps reveals key information about the interaction between the ionized region and the molecular cloud boundary. These boundaries are crucial for delineating the zones where molecular gas is dissociated by ultraviolet radiation from young stars. This analysis allowed us to identify different neutral gas structures, such as filaments and globules, which may trigger new episodes of star formation. In addition, we observed that at the interface between ionized and neutral gas, We suggest that the molecular gas is being compressed as a result of the expansion of the stellar bubble and the evolution of the \Hii\ region.

The radial distribution of emission line ratios enabled the analysis of variations in ionization and temperature within the nebula, providing an estimate of their range of variation. The observed trends in \nii/\Halpha\ and  \sii/\Halpha\ indicate an ionization gradient that could be influenced by variations in the radiation field and/or the density of the medium.

We examine the interaction between young stellar clusters and the nebular gas by identifying stellar populations embedded in the Rosette Nebula. In particular, we explore the presence of stellar clusters such as NGC 2237 in regions with higher concentrations of molecular gas and dust suggests an evolutionary link between the expansion of the ionized region and ongoing star formation activity. Additionally, we analyze some areas containing B-type stars, which could contribute to the evolution of the \Hii\ region through their radiative and dynamic feedback, albeit to a lesser extent.

The obtained results support the hypothesis that the Rosette Nebula formed from a non-homogeneous molecular cloud located at the edge of a filament, exhibiting a thin-sheet structure. This agrees with the model proposed by \cite{Wareing2018}, who suggest that the morphology is shaped by magnetic fields, leading to a sheet-like geometry. A clear correlation is observed between the distribution of molecular gas and dust density with the presence of ionized structures. Furthermore, the data indicate that the expansion of ionized gas has compressed the molecular cloud in certain regions, promoting the formation of dense gas and dust structures. Given that the NGC 2244 cluster has an estimated age of approximately 2 Myr, the results suggest that the ionization, expansion, and gas expulsion processes in lower-density regions occur on a timescale comparable to the lifetime of the cluster.

This study lays the foundation for future kinematic analyses of the Rosette Nebula. A detailed study of the gas velocity structure will help improve our understanding of the dynamic processes shaping the nebula. In a close following study, we will focus on combining optical emission data with kinematic and spectroscopic studies to provide a more comprehensive view of the physical conditions in the Rosette Nebula and their implications for star formation.

\section*{Acknowledgements}

C. R-Z and M. V-D acknowledges support from project UNAM DGAPA-PAPIIT IG 101723, Mexico and project CONAHCYT Ciencia Frontera 86372 (entitled Citlalcoatl). 

J. B-B and M. V-D acknowledge support from project UNAM DGAPA-PAPIIT AG 101025, Mexico. J. B-B thanks support from the PASPA 2025 grant.

M. V-D. was supported by Consejo Nacional de Humanidades, Ciencias y Tecnologías (CONAHCyT) Beca Nacional de Posgrado (pres. Secretaría de Ciencia, Humanidades, Tecnología e Innovación, Secihti )

This research made use of the data from the Milky Way Imaging Scroll Painting (MWISP) project, which is a multi-line survey in 12CO/13CO/C18O along the northern galactic plane with PMO-13.7m telescope. We are grateful to all the members of the MWISP working group, particularly the staff members at PMO-13.7m telescope, for their long-term support. MWISP was sponsored by National Key R\&D Program of China with grant 2017YFA0402701 and CAS Key Research Program of Frontier Sciences with grant QYZDJ-SSW-SLH047.

This research has made use of data products from Herschel Space Telescope, Herschel is an ESA space observatory with science instruments provided by European-led Principal Investigator consortia and with important participation from NASA

Funding for the Sloan Digital Sky Survey V has been provided by the Alfred P. Sloan Foundation, the Heising-Simons Foundation, the National Science Foundation, and the Participating Institutions. SDSS acknowledges support and resources from the Center for High-Performance Computing at the University of Utah. SDSS telescopes are located at Apache Point Observatory, funded by the Astrophysical Research Consortium and operated by New Mexico State University, and at Las Campanas Observatory, operated by the Carnegie Institution for Science. The SDSS web site is \url{www.sdss.org}.

SDSS is managed by the Astrophysical Research Consortium for the Participating Institutions of the SDSS Collaboration, including the Carnegie Institution for Science, Chilean National Time Allocation Committee (CNTAC) ratified researchers, Caltech, the Gotham Participation Group, Harvard University, Heidelberg University, The Flatiron Institute, The Johns Hopkins University, L'Ecole polytechnique f\'{e}d\'{e}rale de Lausanne (EPFL), Leibniz-Institut f\"{u}r Astrophysik Potsdam (AIP), Max-Planck-Institut f\"{u}r Astronomie (MPIA Heidelberg), Max-Planck-Institut f\"{u}r Extraterrestrische Physik (MPE), Nanjing University, National Astronomical Observatories of China (NAOC), New Mexico State University, The Ohio State University, Pennsylvania State University, Smithsonian Astrophysical Observatory, Space Telescope Science Institute (STScI), the Stellar Astrophysics Participation Group, Universidad Nacional Aut\'{o}noma de M\'{e}xico, University of Arizona, University of Colorado Boulder, University of Illinois at Urbana-Champaign, University of Toronto, University of Utah, University of Virginia, Yale University, and Yunnan University

This research made use of Montage. It is funded by the National Science Foundation under Grant Number ACI-1440620, and was previously funded by the National Aeronautics and Space Administration's Earth Science Technology Office, Computation Technologies Project, under Cooperative Agreement Number NCC5-626 between NASA and the California Institute of Technology.

S.F.S. thanks the support by UNAM PASPA - DGAPA  and SECIHTI CBF-2025-I236 project. Authors acknowledge financial support from Spanish Ministry of Science and Innovation (MICINN), project PID2019-107408GB-C43 (ESTALLIDOS)

J.G.F-T gratefully acknowledges the grants support provided by ANID Fondecyt Postdoc No. 3230001 (Sponsoring researcher), and from the Joint Committee ESO-Government of Chile under the agreement 2023 ORP 062/2023.

E.J.J. gratefully acknowledges support by the ANID BASAL project FB210003 and the Millennium Nucleus ERIS NCN2021\_017, Centros ANID Iniciativa Milenio.

K. K. gratefully acknowledges funding from the Deutsche Forschungsgemeinschaft (DFG, German Research Foundation) in the form of an Emmy Noether Research Group (grant number KR4598/2-1, PI Kreckel) and the European Research Council’s starting grant ERC StG-101077573 ("ISM-METALS").

A. S. gratefully acknowledges support by the Fondecyt Regular (project code 1220610), and ANID BASAL project FB210003.

G.A.B. acknowledges the support from the ANID Basal project FB210003. 

Z.L.A. gratefully acknowledges the support provided by the Postdoctoral Program (POSDOC) of UNAM (Universidad Nacional Autónoma de México)

J. E. M-D thanks the support by SECIHTI CBF-2025-I-2048 project ``Resolviendo la Física Interna de las Galaxias: De las Escalas Locales a la Estructura Global con el SDSS-V Local Volume Mapper''. 

\bibliographystyle{mnras}
\bibliography{referencias} 

\begin{thebibliography}{}
\makeatletter
\relax
\def\mn@urlcharsother{\let\do\@makeother \do\$\do\&\do\#\do\^\do\_\do\%\do\~}
\def\mn@doi{\begingroup\mn@urlcharsother \@ifnextchar [ {\mn@doi@} {\mn@doi@[]}}
\def\mn@doi@[#1]#2{\def\@tempa{#1}\ifx\@tempa\@empty \href {http://dx.doi.org/#2} {doi:#2}\else \href {http://dx.doi.org/#2} {#1}\fi \endgroup}
\def\mn@eprint#1#2{\mn@eprint@#1:#2::\@nil}
\def\mn@eprint@arXiv#1{\href {http://arxiv.org/abs/#1} {{\tt arXiv:#1}}}
\def\mn@eprint@dblp#1{\href {http://dblp.uni-trier.de/rec/bibtex/#1.xml} {dblp:#1}}
\def\mn@eprint@#1:#2:#3:#4\@nil{\def\@tempa {#1}\def\@tempb {#2}\def\@tempc {#3}\ifx \@tempc \@empty \let \@tempc \@tempb \let \@tempb \@tempa \fi \ifx \@tempb \@empty \def\@tempb {arXiv}\fi \@ifundefined {mn@eprint@\@tempb}{\@tempb:\@tempc}{\expandafter \expandafter \csname mn@eprint@\@tempb\endcsname \expandafter{\@tempc}}}

\bibitem[\protect\citeauthoryear{{Bagnulo}, {Hensberge}, {Landstreet}, {Szeifert}  \& {Wade}}{{Bagnulo} et~al.}{2004}]{Bagnulo2004}
{Bagnulo} S.,  {Hensberge} H.,  {Landstreet} J.~D.,  {Szeifert} T.,   {Wade} G.~A.,  2004, \mn@doi [\aap] {10.1051/0004-6361:20034283}, \href {https://ui.adsabs.harvard.edu/abs/2004A&A...416.1149B} {416, 1149}

\bibitem[\protect\citeauthoryear{{Barrera-Ballesteros} et~al.,}{{Barrera-Ballesteros} et~al.}{2021}]{Barrera2021}
{Barrera-Ballesteros} J.~K.,  et~al., 2021, \mn@doi [MNRAS] {10.1093/mnras/stab755}, \href {https://ui.adsabs.harvard.edu/abs/2021MNRAS.503.3643B} {503, 3643}

\bibitem[\protect\citeauthoryear{{Bruhweiler}, {Freire Ferrero}, {Bourdin}  \& {Gull}}{{Bruhweiler} et~al.}{2010}]{Bruhweiler2010}
{Bruhweiler} F.~C.,  {Freire Ferrero} R.,  {Bourdin} M.~O.,   {Gull} T.~R.,  2010, \mn@doi [\apj] {10.1088/0004-637X/719/2/1872}, \href {https://ui.adsabs.harvard.edu/abs/2010ApJ...719.1872B} {719, 1872}

\bibitem[\protect\citeauthoryear{{Cambr{\'e}sy}, {Marton}, {Feher}, {T{\'o}th}  \& {Schneider}}{{Cambr{\'e}sy} et~al.}{2013}]{Cambresy2013}
{Cambr{\'e}sy} L.,  {Marton} G.,  {Feher} O.,  {T{\'o}th} L.~V.,   {Schneider} N.,  2013, \mn@doi [\aap] {10.1051/0004-6361/201321235}, \href {https://ui.adsabs.harvard.edu/abs/2013A&A...557A..29C} {557, A29}

\bibitem[\protect\citeauthoryear{{Celnik}}{{Celnik}}{1985}]{Celnik1985}
{Celnik} W.~E.,  1985, \aap, \href {https://ui.adsabs.harvard.edu/abs/1985A&A...144..171C} {144, 171}

\bibitem[\protect\citeauthoryear{{Dent} et~al.,}{{Dent} et~al.}{2009}]{Dent2009}
{Dent} W.~R.~F.,  et~al., 2009, \mn@doi [\mnras] {10.1111/j.1365-2966.2009.14678.x}, \href {https://ui.adsabs.harvard.edu/abs/2009MNRAS.395.1805D} {395, 1805}

\bibitem[\protect\citeauthoryear{{Drory} et~al.,}{{Drory} et~al.}{2024}]{Niv2024}
{Drory} N.,  et~al., 2024, \mn@doi [\aj] {10.3847/1538-3881/ad6de9}, \href {https://ui.adsabs.harvard.edu/abs/2024AJ....168..198D} {168, 198}

\bibitem[\protect\citeauthoryear{{Gahm}, {Carlqvist}, {Johansson}  \& {Nikoli{\'c}}}{{Gahm} et~al.}{2006}]{Gahm2006}
{Gahm} G.~F.,  {Carlqvist} P.,  {Johansson} L.~E.~B.,   {Nikoli{\'c}} S.,  2006, \mn@doi [\aap] {10.1051/0004-6361:20054494}, \href {https://ui.adsabs.harvard.edu/abs/2006A&A...454..201G} {454, 201}

\bibitem[\protect\citeauthoryear{{Gahm}, {Grenman}, {Fredriksson}  \& {Kristen}}{{Gahm} et~al.}{2007}]{Gahm2007}
{Gahm} G.~F.,  {Grenman} T.,  {Fredriksson} S.,   {Kristen} H.,  2007, \mn@doi [\aj] {10.1086/512036}, \href {https://ui.adsabs.harvard.edu/abs/2007AJ....133.1795G} {133, 1795}

\bibitem[\protect\citeauthoryear{{Gaia Collaboration} et~al.,}{{Gaia Collaboration} et~al.}{2021}]{BailerJones2021}
{Gaia Collaboration} et~al., 2021, \mn@doi [\aap] {10.1051/0004-6361/202039657}, \href {https://ui.adsabs.harvard.edu/abs/2021A&A...649A...1G} {649, A1}

\bibitem[\protect\citeauthoryear{{Griffin} et~al.,}{{Griffin} et~al.}{2010}]{Griffin2010}
{Griffin} M.~J.,  et~al., 2010, \mn@doi [A\&A] {10.1051/0004-6361/201014519}, \href {https://ui.adsabs.harvard.edu/abs/2010A&A...518L...3G} {518, L3}

\bibitem[\protect\citeauthoryear{{Hennemann} et~al.,}{{Hennemann} et~al.}{2010}]{Hennemann10}
{Hennemann} M.,  et~al., 2010, \mn@doi [\aap] {10.1051/0004-6361/201014629}, \href {https://ui.adsabs.harvard.edu/abs/2010A&A...518L..84H} {518, L84}

\bibitem[\protect\citeauthoryear{{Herbst} et~al.,}{{Herbst} et~al.}{2024}]{Herbst2024}
{Herbst} T.~M.,  et~al., 2024, \mn@doi [\aj] {10.3847/1538-3881/ad794810.1134/S1063772908070044}, \href {https://ui.adsabs.harvard.edu/abs/2024AJ....168..267H} {168, 267}

\bibitem[\protect\citeauthoryear{{Hopkins}, {Kere{\v{s}}}, {O{\~n}orbe}, {Faucher-Gigu{\`e}re}, {Quataert}, {Murray}  \& {Bullock}}{{Hopkins} et~al.}{2014}]{Hopkins2014}
{Hopkins} P.~F.,  {Kere{\v{s}}} D.,  {O{\~n}orbe} J.,  {Faucher-Gigu{\`e}re} C.-A.,  {Quataert} E.,  {Murray} N.,   {Bullock} J.~S.,  2014, \mn@doi [MNRAS] {10.1093/mnras/stu1738}, \href {https://ui.adsabs.harvard.edu/abs/2014MNRAS.445..581H} {445, 581}

\bibitem[\protect\citeauthoryear{{Hu}, {Lazarian}  \& {Stanimirovi{\'c}}}{{Hu} et~al.}{2021}]{HuYue21}
{Hu} Y.,  {Lazarian} A.,   {Stanimirovi{\'c}} S.,  2021, \mn@doi [\apj] {10.3847/1538-4357/abedb7}, \href {https://ui.adsabs.harvard.edu/abs/2021ApJ...912....2H} {912, 2}

\bibitem[\protect\citeauthoryear{{Kewley}, {Nicholls}  \& {Sutherland}}{{Kewley} et~al.}{2019}]{Kewley2019}
{Kewley} L.~J.,  {Nicholls} D.~C.,   {Sutherland} R.~S.,  2019, \mn@doi [\araa] {10.1146/annurev-astro-081817-051832}, \href {https://ui.adsabs.harvard.edu/abs/2019ARA&A..57..511K} {57, 511}

\bibitem[\protect\citeauthoryear{{Kim}, {Kim}  \& {Ostriker}}{{Kim} et~al.}{2018}]{Kim2018}
{Kim} J.-G.,  {Kim} W.-T.,   {Ostriker} E.~C.,  2018, \mn@doi [APJ] {10.3847/1538-4357/aabe27}, \href {https://ui.adsabs.harvard.edu/abs/2018ApJ...859...68K} {859, 68}

\bibitem[\protect\citeauthoryear{Kollmeier et~al.,}{Kollmeier et~al.}{2025}]{Kollmeier2025}
Kollmeier J.~A.,  et~al., 2025, Sloan Digital Sky Survey-V: Pioneering Panoptic Spectroscopy (\mn@eprint {arXiv} {2507.06989}), \url {https://arxiv.org/abs/2507.06989}

\bibitem[\protect\citeauthoryear{{Konidaris} et~al.,}{{Konidaris} et~al.}{2024}]{Konidaris2024}
{Konidaris} N.~P.,  et~al., 2024, in {Bryant} J.~J.,  {Motohara} K.,   {Vernet} J. R.~D.,  eds,  Society of Photo-Optical Instrumentation Engineers (SPIE) Conference Series Vol. 13096, Ground-based and Airborne Instrumentation for Astronomy X. p. 130961Z, \mn@doi{10.1117/12.3019892}

\bibitem[\protect\citeauthoryear{{Kreckel} et~al.,}{{Kreckel} et~al.}{2024}]{Kreckel2024}
{Kreckel} K.,  et~al., 2024, \mn@doi [\aap] {10.1051/0004-6361/202449943}, \href {https://ui.adsabs.harvard.edu/abs/2024A&A...689A.352K} {689, A352}

\bibitem[\protect\citeauthoryear{{Krumholz} et~al.,}{{Krumholz} et~al.}{2014}]{feedback2014}
{Krumholz} M.~R.,  et~al., 2014, in {Beuther} H.,  {Klessen} R.~S.,  {Dullemond} C.~P.,   {Henning} T.,  eds, Protostars and Planets VI. pp 243--266 (\mn@eprint {arXiv} {1401.2473}), \mn@doi{10.2458/azu_uapress_9780816531240-ch011}

\bibitem[\protect\citeauthoryear{{Li}, {Wang}, {Zhang}, {Ma}, {Fang}  \& {Yang}}{{Li} et~al.}{2018}]{mwisp2018}
{Li} C.,  {Wang} H.,  {Zhang} M.,  {Ma} Y.,  {Fang} M.,   {Yang} J.,  2018, \mn@doi [\apjs] {10.3847/1538-4365/aad963}, \href {https://ui.adsabs.harvard.edu/abs/2018ApJS..238...10L} {238, 10}

\bibitem[\protect\citeauthoryear{{Lim} et~al.,}{{Lim} et~al.}{2021}]{Lim2021}
{Lim} B.,  et~al., 2021, \mn@doi [AJ] {10.3847/1538-3881/abffd8}, \href {https://ui.adsabs.harvard.edu/abs/2021AJ....162...56L} {162, 56}

\bibitem[\protect\citeauthoryear{{Lombardi}, {Bouy}, {Alves}  \& {Lada}}{{Lombardi} et~al.}{2014}]{Lombardi+14}
{Lombardi} M.,  {Bouy} H.,  {Alves} J.,   {Lada} C.~J.,  2014, \mn@doi [\aap] {10.1051/0004-6361/201323293}, \href {https://ui.adsabs.harvard.edu/abs/2014A&A...566A..45L} {566, A45}

\bibitem[\protect\citeauthoryear{{Madsen}, {Reynolds}  \& {Haffner}}{{Madsen} et~al.}{2006}]{Madsen2006}
{Madsen} G.~J.,  {Reynolds} R.~J.,   {Haffner} L.~M.,  2006, \mn@doi [\apj] {10.1086/508441}, \href {https://ui.adsabs.harvard.edu/abs/2006ApJ...652..401M} {652, 401}

\bibitem[\protect\citeauthoryear{{Mahy}, {Naz{\'e}}, {Rauw}, {Gosset}, {De Becker}, {Sana}  \& {Eenens}}{{Mahy} et~al.}{2009}]{Mahy2009}
{Mahy} L.,  {Naz{\'e}} Y.,  {Rauw} G.,  {Gosset} E.,  {De Becker} M.,  {Sana} H.,   {Eenens} P.,  2009, \mn@doi [\aap] {10.1051/0004-6361/200911662}, \href {https://ui.adsabs.harvard.edu/abs/2009A&A...502..937M} {502, 937}

\bibitem[\protect\citeauthoryear{{M{\"a}kel{\"a}}, {Haikala}  \& {Gahm}}{{M{\"a}kel{\"a}} et~al.}{2014}]{Makela2014}
{M{\"a}kel{\"a}} M.~M.,  {Haikala} L.~K.,   {Gahm} G.~F.,  2014, \mn@doi [\aap] {10.1051/0004-6361/201423440}, \href {https://ui.adsabs.harvard.edu/abs/2014A&A...567A.108M} {567, A108}

\bibitem[\protect\citeauthoryear{{M{\"a}kel{\"a}}, {Haikala}  \& {Gahm}}{{M{\"a}kel{\"a}} et~al.}{2017}]{Makela2017}
{M{\"a}kel{\"a}} M.~M.,  {Haikala} L.~K.,   {Gahm} G.~F.,  2017, \mn@doi [\aap] {10.1051/0004-6361/201525655}, \href {https://ui.adsabs.harvard.edu/abs/2017A&A...605A..82M} {605, A82}

\bibitem[\protect\citeauthoryear{{Martins}, {Mahy}, {Hillier}  \& {Rauw}}{{Martins} et~al.}{2012}]{Martins2012}
{Martins} F.,  {Mahy} L.,  {Hillier} D.~J.,   {Rauw} G.,  2012, \mn@doi [\aap] {10.1051/0004-6361/201117458}, \href {https://ui.adsabs.harvard.edu/abs/2012A&A...538A..39M} {538, A39}

\bibitem[\protect\citeauthoryear{{McLeod}, {Dale}, {Ginsburg}, {Ercolano}, {Gritschneder}, {Ramsay}  \& {Testi}}{{McLeod} et~al.}{2015}]{McLeod2015}
{McLeod} A.~F.,  {Dale} J.~E.,  {Ginsburg} A.,  {Ercolano} B.,  {Gritschneder} M.,  {Ramsay} S.,   {Testi} L.,  2015, \mn@doi [\mnras] {10.1093/mnras/stv680}, \href {https://ui.adsabs.harvard.edu/abs/2015MNRAS.450.1057M} {450, 1057}

\bibitem[\protect\citeauthoryear{{McLeod}, {Dale}, {Evans}, {Ginsburg}, {Kruijssen}, {Pellegrini}, {Ramsay}  \& {Testi}}{{McLeod} et~al.}{2019}]{McLeod2019}
{McLeod} A.~F.,  {Dale} J.~E.,  {Evans} C.~J.,  {Ginsburg} A.,  {Kruijssen} J.~M.~D.,  {Pellegrini} E.~W.,  {Ramsay} S.~K.,   {Testi} L.,  2019, \mn@doi [MNRAS] {10.1093/mnras/sty2696}, \href {https://ui.adsabs.harvard.edu/abs/2019MNRAS.486.5263M} {486, 5263}

\bibitem[\protect\citeauthoryear{{Meisner} \& {Finkbeiner}}{{Meisner} \& {Finkbeiner}}{2014}]{meisner2014}
{Meisner} A.~M.,  {Finkbeiner} D.~P.,  2014, \mn@doi [\apj] {10.1088/0004-637X/781/1/5}, \href {https://ui.adsabs.harvard.edu/abs/2014ApJ...781....5M} {781, 5}

\bibitem[\protect\citeauthoryear{{Mejia}}{{Mejia}}{prep}]{mejia24}
{Mejia} A.,  in prep, {LVM Data Reduction Pipeline}

\bibitem[\protect\citeauthoryear{{Mellema}, {Arthur}, {Henney}, {Iliev}  \& {Shapiro}}{{Mellema} et~al.}{2006}]{Mellema06}
{Mellema} G.,  {Arthur} S.~J.,  {Henney} W.~J.,  {Iliev} I.~T.,   {Shapiro} P.~R.,  2006, \mn@doi [\apj] {10.1086/505294}, \href {https://ui.adsabs.harvard.edu/abs/2006ApJ...647..397M} {647, 397}

\bibitem[\protect\citeauthoryear{{Moran}, {Ybarra}  \& {Rom{\'a}n-Z{\'u}{\~n}iga}}{{Moran} et~al.}{2024}]{Moran+2024}
{Moran} E.,  {Ybarra} J.~E.,   {Rom{\'a}n-Z{\'u}{\~n}iga} C.,  2024, in American Astronomical Society Meeting Abstracts. p. 402.19

\bibitem[\protect\citeauthoryear{{Motte} et~al.,}{{Motte} et~al.}{2010}]{Motte2010}
{Motte} F.,  et~al., 2010, \mn@doi [\aap] {10.1051/0004-6361/201014690}, \href {https://ui.adsabs.harvard.edu/abs/2010A&A...518L..77M} {518, L77}

\bibitem[\protect\citeauthoryear{{Mu{\v{z}}i{\'c}}, {Scholz}, {Pe{\~n}a Ram{\'\i}rez}, {Jayawardhana}, {Sch{\"o}del}, {Geers}, {Cieza}  \& {Bayo}}{{Mu{\v{z}}i{\'c}} et~al.}{2019}]{Muzic2019}
{Mu{\v{z}}i{\'c}} K.,  {Scholz} A.,  {Pe{\~n}a Ram{\'\i}rez} K.,  {Jayawardhana} R.,  {Sch{\"o}del} R.,  {Geers} V.~C.,  {Cieza} L.~A.,   {Bayo} A.,  2019, \mn@doi [\apj] {10.3847/1538-4357/ab2da4}, \href {https://ui.adsabs.harvard.edu/abs/2019ApJ...881...79M} {881, 79}

\bibitem[\protect\citeauthoryear{{Mu{\v{z}}i{\'c}}, {Almendros-Abad}, {Bouy}, {Kubiak}, {Pe{\~n}a Ram{\'\i}rez}, {Krone-Martins}, {Moitinho}  \& {Concei{\c{c}}{\~a}o}}{{Mu{\v{z}}i{\'c}} et~al.}{2022}]{Muzic2022}
{Mu{\v{z}}i{\'c}} K.,  {Almendros-Abad} V.,  {Bouy} H.,  {Kubiak} K.,  {Pe{\~n}a Ram{\'\i}rez} K.,  {Krone-Martins} A.,  {Moitinho} A.,   {Concei{\c{c}}{\~a}o} M.,  2022, \mn@doi [\aap] {10.1051/0004-6361/202243659}, \href {https://ui.adsabs.harvard.edu/abs/2022A&A...668A..19M} {668, A19}

\bibitem[\protect\citeauthoryear{Perruchot et~al.,}{Perruchot et~al.}{2018}]{Perruchot2018}
Perruchot S.,  et~al., 2018, in Evans C.~J.,  Simard L.,   Takami H.,  eds,  Proceedings of SPIE Vol. 10702, Ground-based and Airborne Instrumentation for Astronomy VII. SPIE, p. 107027K, \mn@doi{10.1117/12.2311996}, \url {https://doi.org/10.1117/12.2311996}

\bibitem[\protect\citeauthoryear{{Phelps} \& {Lada}}{{Phelps} \& {Lada}}{1997}]{PL97}
{Phelps} R.~L.,  {Lada} E.~A.,  1997, \mn@doi [\apj] {10.1086/303713}, \href {https://ui.adsabs.harvard.edu/abs/1997ApJ...477..176P} {477, 176}

\bibitem[\protect\citeauthoryear{{Poglitsch} et~al.,}{{Poglitsch} et~al.}{2010}]{Poglitsch2010}
{Poglitsch} A.,  et~al., 2010, \mn@doi [\aap] {10.1051/0004-6361/201014535}, \href {https://ui.adsabs.harvard.edu/abs/2010A&A...518L...2P} {518, L2}

\bibitem[\protect\citeauthoryear{{Poulton}, {Robitaille}, {Greaves}, {Bonnell}, {Williams}  \& {Heyer}}{{Poulton} et~al.}{2008}]{Poulton08}
{Poulton} C.~J.,  {Robitaille} T.~P.,  {Greaves} J.~S.,  {Bonnell} I.~A.,  {Williams} J.~P.,   {Heyer} M.~H.,  2008, \mn@doi [\mnras] {10.1111/j.1365-2966.2007.12556.x}, \href {https://ui.adsabs.harvard.edu/abs/2008MNRAS.384.1249P} {384, 1249}

\bibitem[\protect\citeauthoryear{{Pribulla} et~al.,}{{Pribulla} et~al.}{2010}]{Pribulla2010}
{Pribulla} T.,  et~al., 2010, \mn@doi [Astronomische Nachrichten] {10.1002/asna.201011351}, \href {https://ui.adsabs.harvard.edu/abs/2010AN....331..397P} {331, 397}

\bibitem[\protect\citeauthoryear{{Rom{\'a}n-Z{\'u}{\~n}iga} \& {Lada}}{{Rom{\'a}n-Z{\'u}{\~n}iga} \& {Lada}}{2008}]{ReviewRoman2008}
{Rom{\'a}n-Z{\'u}{\~n}iga} C.~G.,  {Lada} E.~A.,  2008, in {Reipurth} B.,  ed., , Vol.~4, Handbook of Star Forming Regions, Volume I.
p.~928, \mn@doi{10.48550/arXiv.0810.0931}

\bibitem[\protect\citeauthoryear{{Rom{\'a}n-Z{\'u}{\~n}iga}, {Elston}, {Ferreira}  \& {Lada}}{{Rom{\'a}n-Z{\'u}{\~n}iga} et~al.}{2008}]{Roman2008}
{Rom{\'a}n-Z{\'u}{\~n}iga} C.~G.,  {Elston} R.,  {Ferreira} B.,   {Lada} E.~A.,  2008, \mn@doi [\apj] {10.1086/523785}, \href {https://ui.adsabs.harvard.edu/abs/2008ApJ...672..861R} {672, 861}

\bibitem[\protect\citeauthoryear{{S{\'a}nchez}}{{S{\'a}nchez}}{2020}]{Sanchez2020}
{S{\'a}nchez} S.~F.,  2020, \mn@doi [\araa] {10.1146/annurev-astro-012120-013326}, \href {https://ui.adsabs.harvard.edu/abs/2020ARA&A..58...99S} {58, 99}

\bibitem[\protect\citeauthoryear{{S{\'a}nchez} et~al.,}{{S{\'a}nchez} et~al.}{2025}]{Sanchez2025}
{S{\'a}nchez} S.~F.,  et~al., 2025, \mn@doi [\aj] {10.3847/1538-3881/ad93bb}, \href {https://ui.adsabs.harvard.edu/abs/2025AJ....169...52S} {169, 52}

\bibitem[\protect\citeauthoryear{{Scheuermann} et~al.,}{{Scheuermann} et~al.}{2023}]{Fabian2023}
{Scheuermann} F.,  et~al., 2023, \mn@doi [MNRAS] {10.1093/mnras/stad878}, \href {https://ui.adsabs.harvard.edu/abs/2023MNRAS.522.2369S} {522, 2369}

\bibitem[\protect\citeauthoryear{{Schneider} et~al.,}{{Schneider} et~al.}{2010}]{Schneider10}
{Schneider} N.,  et~al., 2010, \mn@doi [\aap] {10.1051/0004-6361/201014627}, \href {https://ui.adsabs.harvard.edu/abs/2010A&A...518L..83S} {518, L83}

\bibitem[\protect\citeauthoryear{{Shang} et~al.,}{{Shang} et~al.}{2022}]{Shang2022}
{Shang} L.-H.,  et~al., 2022, \mn@doi [\apjs] {10.3847/1538-4365/ac5831}, \href {https://ui.adsabs.harvard.edu/abs/2022ApJS..259...63S} {259, 63}

\bibitem[\protect\citeauthoryear{{Swinyard} et~al.,}{{Swinyard} et~al.}{2010}]{Swinyard2010}
{Swinyard} B.~M.,  et~al., 2010, \mn@doi [\aap] {10.1051/0004-6361/201014605}, \href {https://ui.adsabs.harvard.edu/abs/2010A&A...518L...4S} {518, L4}

\bibitem[\protect\citeauthoryear{{Wang}, {Townsley}, {Feigelson}, {Broos}, {Getman}, {Rom{\'a}n-Z{\'u}{\~n}iga}  \& {Lada}}{{Wang} et~al.}{2008}]{Wang2008}
{Wang} J.,  {Townsley} L.~K.,  {Feigelson} E.~D.,  {Broos} P.~S.,  {Getman} K.~V.,  {Rom{\'a}n-Z{\'u}{\~n}iga} C.~G.,   {Lada} E.,  2008, \mn@doi [\apj] {10.1086/526406}, \href {https://ui.adsabs.harvard.edu/abs/2008ApJ...675..464W} {675, 464}

\bibitem[\protect\citeauthoryear{{Wang}, {Feigelson}, {Townsley}, {Broos}, {Rom{\'a}n-Z{\'u}{\~n}iga}, {Lada}  \& {Garmire}}{{Wang} et~al.}{2010}]{Wang2010}
{Wang} J.,  {Feigelson} E.~D.,  {Townsley} L.~K.,  {Broos} P.~S.,  {Rom{\'a}n-Z{\'u}{\~n}iga} C.~G.,  {Lada} E.,   {Garmire} G.,  2010, \mn@doi [\apj] {10.1088/0004-637X/716/1/474}, \href {https://ui.adsabs.harvard.edu/abs/2010ApJ...716..474W} {716, 474}

\bibitem[\protect\citeauthoryear{{Wareing}, {Pittard}, {Wright}  \& {Falle}}{{Wareing} et~al.}{2018}]{Wareing2018}
{Wareing} C.~J.,  {Pittard} J.~M.,  {Wright} N.~J.,   {Falle} S.~A.~E.~G.,  2018, \mn@doi [\mnras] {10.1093/mnras/sty148}, \href {https://ui.adsabs.harvard.edu/abs/2018MNRAS.475.3598W} {475, 3598}

\bibitem[\protect\citeauthoryear{{Wilson}, {Bania}  \& {Balser}}{{Wilson} et~al.}{2015}]{Wilson2015}
{Wilson} T.~L.,  {Bania} T.~M.,   {Balser} D.~S.,  2015, \mn@doi [\apj] {10.1088/0004-637X/812/1/45}, \href {https://ui.adsabs.harvard.edu/abs/2015ApJ...812...45W} {812, 45}

\bibitem[\protect\citeauthoryear{{Wright} et~al.,}{{Wright} et~al.}{2010}]{Wright2010}
{Wright} E.~L.,  et~al., 2010, \mn@doi [\aj] {10.1088/0004-6256/140/6/1868}, \href {https://ui.adsabs.harvard.edu/abs/2010AJ....140.1868W} {140, 1868}

\bibitem[\protect\citeauthoryear{{Ybarra}, {Lada}, {Rom{\'a}n-Z{\'u}{\~n}iga}, {Balog}, {Wang}  \& {Feigelson}}{{Ybarra} et~al.}{2013}]{Ybarra13}
{Ybarra} J.~E.,  {Lada} E.~A.,  {Rom{\'a}n-Z{\'u}{\~n}iga} C.~G.,  {Balog} Z.,  {Wang} J.,   {Feigelson} E.~D.,  2013, \mn@doi [\apj] {10.1088/0004-637X/769/2/140}, \href {https://ui.adsabs.harvard.edu/abs/2013ApJ...769..140Y} {769, 140}

\bibitem[\protect\citeauthoryear{{Zari}, {Lombardi}, {Alves}, {Lada}  \& {Bouy}}{{Zari} et~al.}{2016}]{Zari+16}
{Zari} E.,  {Lombardi} M.,  {Alves} J.,  {Lada} C.~J.,   {Bouy} H.,  2016, \mn@doi [\aap] {10.1051/0004-6361/201526597}, \href {https://ui.adsabs.harvard.edu/abs/2016A&A...587A.106Z} {587, A106}

\makeatother
\end{thebibliography}


\section*{Data Availability}

The LVM survey data will be made publicly available with SDSS Data Release 20. MWISP data can be accessed upon formal request through the MWISP portal. The WISE WSSA \(12 \, \mu\text{m}\) maps and the Herschel images used to derive the column density map are both publicly available. Parameters of the ionizing stellar sources used in this work are available through public catalogs.
 







\appendix

\section{Maps of emission lines}
\label{sec:appendix_A}

\begin{figure*} 
    \centering
    \includegraphics[width=0.8\textwidth]{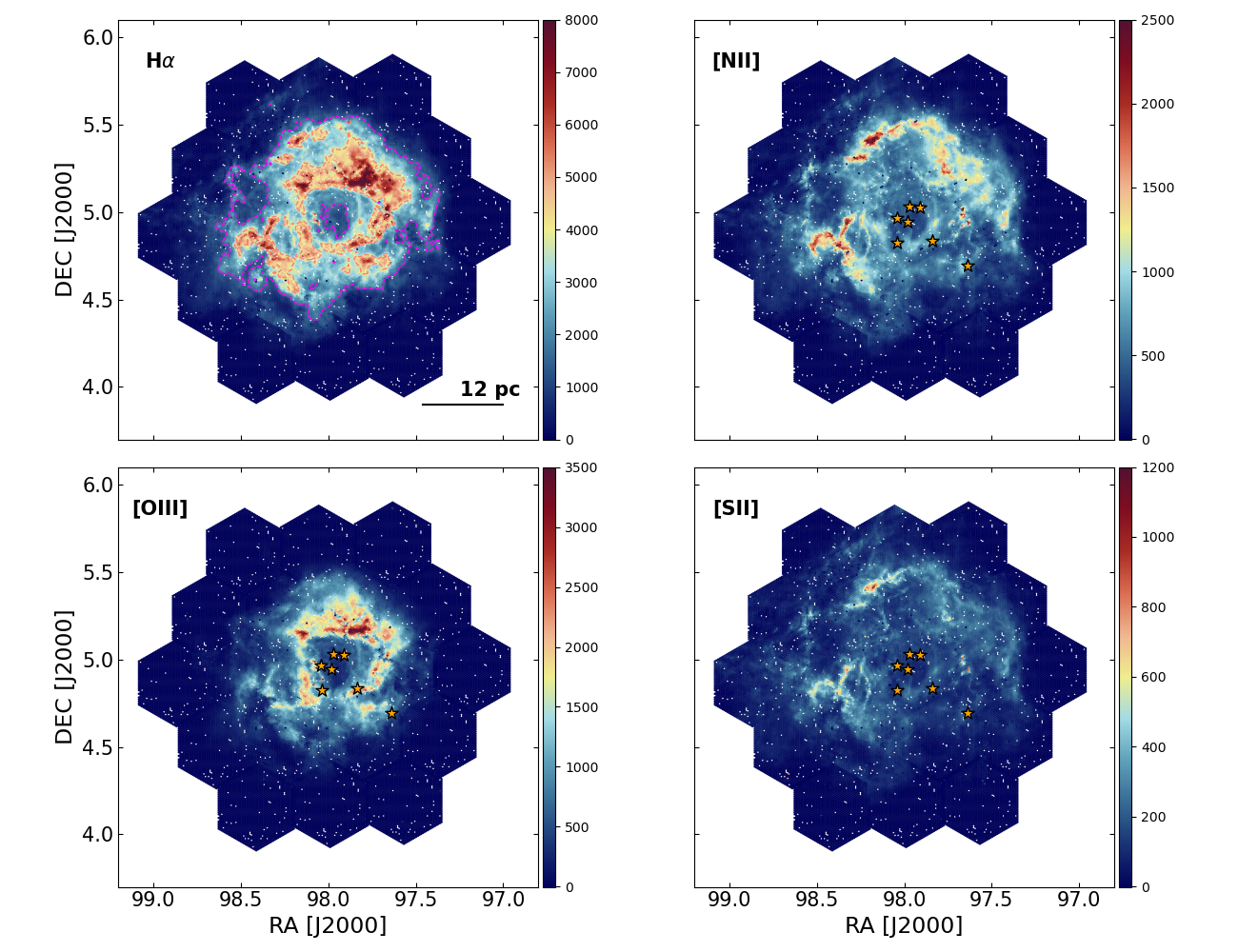} 
    \caption{Relative flux maps of the Rosette Nebula for different emission lines: \Halpha\ (top-left), \nii\ (top-right), \oiii\ (bottom-left), and \sii\ (bottom-right). The \Halpha\ map shows contours at 20\% (magenta) and 70\% (salmon) of the maximum flux for this line, highlighting the structure of the ionized gas. O-type stars are marked with yellow stars. Color bars indicate relative flux intensity. The scale bar in the \Halpha\ panel represents 12 pc, assuming a distance of 1.5 kpc to the nebula.}
    \label{fig:emision_lines_lvm}
\end{figure*}

Figure \ref{fig:emision_lines_lvm} shows the spatial distribution of the relative flux of the four brightest emission lines within the spectral range covered by the LVM: \Halpha, \nii\ $\lambda$6583, \oiii\ $\lambda$5007, and \sii\ $\lambda$ 6717, $\lambda$ 6731. The \Halpha\ (top left panel) reveals a well-defined ring-like structure with a central evacuated cavity, consistent with stellar feedback from the OB stars in the NGC 2244 cluster, in agreement with previous studies \citep{ReviewRoman2008}. The highest fluxes are found toward the northwest.
The \nii\ map traces the structure at the periphery of the \Hii\ region, following the areas of strongest \Halpha\ emission but with a more filamentary morphology, showing emission peaks in the northeast and southeast quadrants. The \oiii\ emission closely follows the layered structures delineated by \Halpha, with a flux maximum also in the northwestern region. In contrast, the \sii\ map shows generally weaker emission but displays a spatial distribution similar to that of \nii, highlighting compact structures in the northwestern and southwestern quadrants, as well as globules in the eastern region.
While the \Halpha\ and \oiii\ emissions outline the annular structure and regions of higher ionization more clearly, the \nii\ and \sii\ lines exhibit higher intensity in the outer regions of the nebula, suggesting intermediate or low ionization conditions at the nebula's boundaries.

\section{Comparison between \Halpha\ and columnar density}
\label{sec:appendix_B}

\begin{figure} 
    \centering
    \includegraphics[width=0.5\textwidth]{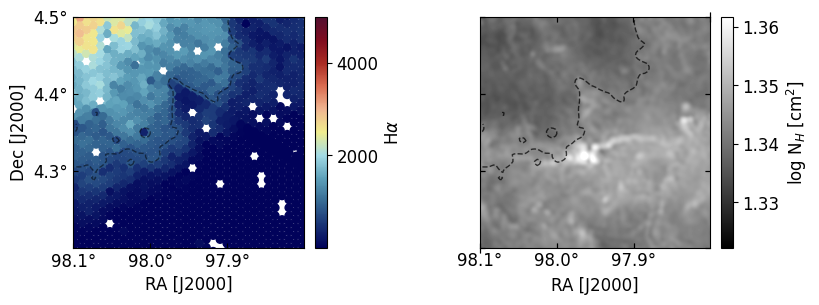} 
    \caption{Low intensity region in the \Halpha emission (left panel), the contour represents 1$\%$ of the maximum relative flux \Halpha\ compared to the column density map (right panel). We highlight how LVM data at low fluxes can reveal detailed structures in the region.}
    \label{fig:A_herschel}
\end{figure}

To explore the detection limits of ionized gas using the Local Volume Mapper (LVM) instrument, we present a comparison between the \Halpha\ emission map and the column density map derived from far-infrared data from Herschel, Figure \ref{fig:A_herschel}. The left panel of Figure B1 shows the faintest emission regions in the \Halpha\ map, where the contour represents 1\% of the relative maximum flux, highlighting the sensitivity of LVM data to low surface brightness structures.  In the right panel, the column density map based on Herschel data illustrates the spatial distribution of dense molecular gas, overlaid with the contour corresponding to low  \Halpha\ flux. A detailed inspection reveals that the ionized gas traced by the faint  \Halpha\ emission contours coincides with regions of high column density. This qualitative anticorrelation supports a scenario in which dense material shields the molecular cloud from ionization and underscores the capability of LVM data to resolve ionization fronts on sub-parsec scales. The comparison demonstrates that low-intensity  \Halpha\ regions can effectively trace the interface between ionized and molecular gas, reinforcing the value of LVM for detailed studies of stellar feedback processes in \Hii\ regions.


\bsp	
\label{lastpage}
\end{document}